\documentclass[onecolumn]{IEEEtran}
\usepackage{cite}

\ifCLASSINFOpdf
\else
\fi
\usepackage{amsmath}
\usepackage{amsthm}
\usepackage{algorithmic}
\ifCLASSOPTIONcompsoc
  \usepackage[caption=false,font=normalsize,labelfont=sf,textfont=sf]{subfig}
\else
  \usepackage[caption=false,font=footnotesize]{subfig}
\fi
\usepackage{amssymb}
\usepackage{xcolor}
\usepackage{graphicx}
\newtheorem{theorem}{Theorem}
\newtheorem{lemma}{Lemma}
\newtheorem{assumption}{Assumption}

\newtheorem{proposition}{Proposition}

\begin{document}
%
\title{Learning Algorithms for \\ Minimizing Queue Length Regret}
%
%
%

\author{\IEEEauthorblockN{Thomas~Stahlbuhk\IEEEauthorrefmark{1}, Brooke~Shrader\IEEEauthorrefmark{1} and~Eytan~Modiano\IEEEauthorrefmark{2}} \\ \thanks{A subset of the material in this paper appeared in \cite{stahlbuhk1}.}
\IEEEauthorblockA{
\IEEEauthorrefmark{1} MIT Lincoln Laboratory, Lexington, MA \\
\IEEEauthorrefmark{2} MIT Laboratory for Information and Decision Systems, Cambridge, MA}
\thanks{This work was sponsored by NSF Grants AST-1547331 and CNS-1701964, and by Army Research Office (ARO) grant number W911NF-17-1-0508. This material is based upon work supported by the United States Air Force under Air Force Contract No. FA8702-15-D-0001. Any opinions, findings, conclusions or recommendations expressed in this material are those of the authors and do not necessarily reflect the views of the United States Air Force.}}

\maketitle

\begin{abstract}
We consider a system consisting of a single transmitter/receiver pair and $N$ channels over which they may communicate.  Packets randomly arrive to the transmitter's queue and wait to be successfully sent to the receiver.  The transmitter may attempt a frame transmission on one channel at a time, where each frame includes a packet if one is in the queue.  For each channel, an attempted transmission is successful with an unknown probability.  The transmitter's objective is to quickly identify the best channel to minimize the number of packets in the queue over $T$ time slots.  To analyze system performance, we introduce queue length regret, which is the expected difference between the total queue length of a learning policy and a controller that knows the rates, a priori.  One approach to designing a transmission policy would be to apply algorithms from the literature that solve the closely-related stochastic multi-armed bandit problem.  These policies would focus on maximizing the number of successful frame transmissions over time.  However, we show that these methods have $\Omega(\log{T})$ queue length regret.  On the other hand, we show that there exists a set of queue-length based policies that can obtain order optimal $O(1)$ queue length regret.  We use our theoretical analysis to devise heuristic methods that are shown to perform well in simulation.
\end{abstract}


%
\IEEEpeerreviewmaketitle

\section{Introduction}
%
%
%
%
\IEEEPARstart{I}{n} this work, we consider a statistical learning problem that is motivated by the following application.  Consider a wireless communication system consisting of a single transmitter/receiver pair and $N$ channels over which they may communicate.  Packets randomly arrive to the transmitter's queue and wait in the queue until they are successfully delivered to the receiver. At each time slot, the transmitter can decide to transmit a frame on one of the $N$ channels.  If the queue is non-empty, the transmitted frame carries a packet with it over the channel, and if the frame is then successfully received, the packet is successfully delivered.  For each channel, each frame transmission attempt is successful according to a probability that is initially unknown. The transmitter is informed whether a transmission was successful via receiver feedback that immediately follows each transmission.  The objective of the controller is to minimize the queue's backlog by using the receiver feedback to quickly identifying the best channel to transmit on.

In the above application, each successful frame transmission offers one packet's worth of service to the queue.  Thus, the channels in the above application behave like servers in a general queueing system that, when selected, offer a random amount of service.  Given the above motivation, we consider the problem of identifying the best of $N$ available servers to minimize the queue's backlog.  To this end, we associate a one unit delay-cost to each time slot that each packet has to wait in the queue.  To obtain good performance, the controller must schedule the servers to explore the offered service rate that each gives and also exploit its knowledge to schedule the server that appears to give the most service.  We define the queue length regret as $R^\pi(T) \triangleq E\left[\sum_{t = 0}^{T-1} Q^\pi(t) - \sum_{t = 0}^{T-1} Q^*(t)\right]$, where $Q^\pi(t)$ is the backlog under a learning policy and $Q^*(t)$ is the backlog under a controller that knows the best server.  Our objective is to find a policy that minimizes this regret.

Our problem is closely related to the stochastic multi-armed bandit problem.  In this problem, a player is confronted by a set of $N$ possible actions of which it may select only one at a given time.  Each action provides an i.i.d. stochastic reward, but the statistics of the rewards are initially unknown.   Over a set of $T$ successive rounds, the player must select from the actions to both explore how much reward each gives as well as exploit its knowledge to focus on the action that appears to give the most reward.  Learning policies for solving the multi-armed bandit have long been considered \cite{thompson}.  Historically, the performance of a policy is evaluated using regret, which is defined to be the expected difference between the reward accumulated by the learning policy and a player that knows, a priori, the action with highest mean reward.  This quantifies the cost of having to learn the best action.  It is well known that there exist policies such that the regret scales on the order of $\log{T}$ and that the order of this bound is tight \cite{bubeck}.  In the seminal work of \cite{lai}, policies for achieving an asymptotically efficient regret rate were derived, and subsequent work in \cite{agarwal, auer, garivier} have provided simplified policies that are often used in practice.

One approach to solving our problem would be to simply use traditional bandit algorithms.  In this approach, the reward in the bandit problem would be viewed as offered service in our problem, and the resulting methods would focus on maximizing the rate of offered service to the queue.  In the context of the wireless system described above, this is equivalent to targeting a high rate of successful frame transmissions without regard to which frames are packet-bearing.  Note that in an infinitely backlogged system, which always has packets available to send, this approach would maximize throughput.  Since, as a rule of thumb, higher service throughputs generally correspond with lower packet delays in queueing systems, this approach has merit.

However, the drawback with this approach is that it does not exploit the fundamental queueing dynamics of the system.  During time periods when the queue is empty, the offered service is unused, and the controller can, therefore, freely poll the servers without hurting its objective. This contrasts with non-empty periods, when exploring can be costly since it potentially uses a suboptimal server.  However, a controller cannot take this lesson too far and restrict its explorations only to time slots when the queue is empty.  This is because some servers may have a service rate that is below the packet arrival rate, and if the controller refuses to explore during non-empty periods, it may settle on destabilizing actions that cause the backlog to grow to infinity.

Instead, policies should favor exploration during periods when the queue is empty but perform enough exploration during non-empty periods to maintain stability.  In this work, we show that for systems that have at least one server whose service rate is greater than the arrival rate there exist queue-length based policies such that $R^\pi(T) = O(1)$.\footnote{For functions $f$ and $g$, $f(x) = O(g(x))$ \emph{iff} $\exists M > 0$ and $\exists x_0$ such that $\forall x > x_0$, $\lvert f(x) \rvert \leq M g(x)$.  Likewise, $f(x) = \Omega(g(x))$ \emph{iff} $\exists M > 0$ such that $\forall x_0$, $\exists x > x_0$ with $\lvert f(x) \rvert \geq M g(x)$.} Likewise, we show that any traditional bandit learning algorithm, when applied to our setting, can have queue length regret $R^\pi(T) = \Omega(\log{T})$.  We additionally show that if there does not exist a server with a service rate that is greater than the arrival rate, for any policy the queue-length regret can grow as $R^\pi(T) = \Omega(T)$.

The problem considered herein is related to the work of \cite{krishnasamy1}, which also considered learning algorithms for scheduling service to a queue. The main difference is that in \cite{krishnasamy1} the controller did not use the queue backlog to make decisions.  As a result, the policies considered in that work were closely related to those in the bandit literature and focused on maximizing offered service.  Under these policies, \cite{krishnasamy1} showed that the tail of $E\left[Q^\pi(t) - Q^*(t)\right]$ diminishes as $\frac{1}{t}$, implying that $R^\pi(T)$ is logarithmic.  In this work, we show that by exploiting empty periods, the queue length regret is bounded.

Our problem can also be compared to the more general literature on reinforcement learning. In reinforcement learning, an actor seeks to learn the optimal policy for a sequential decision making problem with an evolving system state.  The decision made by the actor at each state causes the actor to obtain a probabilistic reward and the system to randomly change its state.  The objective is to learn how to obtain high reward by estimating the system's statistics.  See \cite{bertsekas, sutton} for a review of reinforcement learning.  The online performance of reinforcement learning algorithms has been previously explored in \cite{auer_rl}~and~\cite{jaksch}, which used the upper confidence methods UCRL and UCRL2. The work of \cite{azar} has improved upon these bounds, and \cite{fruit} extended the results to weakly-communicating problems.  Thompson-sampling-inspired algorithms have also emerged as a design principle \cite{russo}. Posterior sampling reinforcement learning (PSRL), which is based on this principle, has become a popular method that can, in certain problems, outperform upper confidence methods \cite{strens, osband1, osband2, osband3}.

Our work can be viewed as a reinforcement learning problem with known structure.  The queue backlog is the problem's state and a penalty is paid for inefficiently scheduling whenever the backlog is nonzero.  Note that in our problem, the amount of regret that can be accrued at a given time is unbounded and the problem's state space (the queue occupancy) is infinite.  Our work has some overlap with the field of safe reinforcement learning \cite{modolvan, berkenkamp, wachi}.  We seek to find learning policies that do not destabilize the system and regularly return the queue backlog to the empty state.  This structure separates our work from most of the previous literature that explores algorithms that are not designed to exploit queueing dynamics and are not focused on analyzing queue stability under different policies.  The algorithms we design obtain their performance guarantees by using what is known about the problem, the dynamics of the queue and its impact on regret, while focusing on learning what is unknown, the server rates. Since the amount of service that a server will offer when scheduled is independent of the current queue backlog, our methods do not waste time trying to match the action-rewards to individual states, as would occur with a general reinforcement learning algorithm that knew nothing about the problem's structure.

The use of statistical learning methods for optimizing wireless channel access in systems with uncertain channel conditions has been considered in the previous literature~\cite{anandkumar, avner, cayci, combes, gai1, gai2, kalathil, lelarge, liu_bandit, nayyar, tekin, zhang, zhou}.  However, most previous works examined algorithms for maximizing transmission opportunities and did not focus on minimizing queueing delay.  A major contribution of this work is to show that these two objectives are not necessarily the same.  A learning algorithm that maximizes total offered service to a queue may not statistically minimize the queue backlog over time.

Scheduling algorithms that use queue state to make service decisions have a long history.  In the seminal work of \cite{tassiulas} and \cite{tassiulas2}, the max-weight algorithm for assigning service to queues was shown to maximize throughput under complex scheduling constraints and probabilistic dynamics.  This framework has been extended and applied to network switching \cite{mckeown}, satellite communications \cite{neely0}, ad-hoc networking \cite{lin, chen2}, packet multicasting and broadcasting \cite{sinha}, packet-delivery-time reduction \cite{joo2}, multi-user MIMO \cite{bethanabhotla}, energy harvesting systems \cite{yu}, and age-of-information minimization \cite{kadota1, kadota2}.
In the works of \cite{krishnasamy2}~and~\cite{stahlbuhk2}, learning algorithms were used for achieving network stability under unknown arrival and channel statistics.  The methods considered in those works augmented the max-weight algorithm with a statistical learning component.  The resulting methods were shown to achieve stability but did not analyze queue length regret.  In \cite{liang}, learning algorithms for controlling networks with adversarial dynamics were considered and the max-weight algorithm and tracking algorithm were both explored as solutions for minimizing queue backlogs.  However, the adversarial model considered in that work is pessimistic compared to the stochastic model considered herein.  

The focus of this work is on proving the achievability of asymptotically-optimal regret growth.  This requires that the controller samples all servers often enough to converge on the optimal server and achieve bounded regret.  Importantly, the controller must concurrently guarantee that the queue does not blow-up to infinity and that exploration of sub-optimal servers is eventually limited to times when the queue is empty.  The policies we use to prove our results do not optimize the queue length regret for finite-time analysis.  As a result, although the tail of their regret is optimal, they can initially accrue large regret.  We designed these policies to facilitate analysis, which is made difficult by the high-correlation between the queue backlog (which determines when to sample the servers to converge on the best server) and the observations of the servers (which determine the queue backlog and are used for identifying the best server).  Our analyzed algorithms disentangle this relationship at the expense of good non-asymptotic performance. Nevertheless, we show through simulation that the insight gained from our analysis can inspire well-performing heuristics that exploit the structure found in our theoretical proofs.  Analyzing and improving upon these heuristics is a potential direction for future research.

The rest of this paper is organized as follows.  In Section~\ref{section_0}, we give the problem setup.  We analyze the performance of traditional bandit algorithms, when applied to our problem, in Section~\ref{section_regret_bandit_alg}.  In Sections~\ref{section_1}~and~\ref{section_1_4}, we characterize the regret of queue-length based policies when the best server has a service rate above and below the arrival rate, respectively. In Section~\ref{section_simulation}, we use our theoretical analysis to define heuristic policies and test their performance through simulation.  We conclude in Section~\ref{conclude}.  Note that a subset of the material in this paper appeared in \cite{stahlbuhk1}.

\section{Problem Setup}
\label{section_0}

\subsection{Problem Description}

\begin{figure}
\begin{center}
    \includegraphics[width=0.5\linewidth]{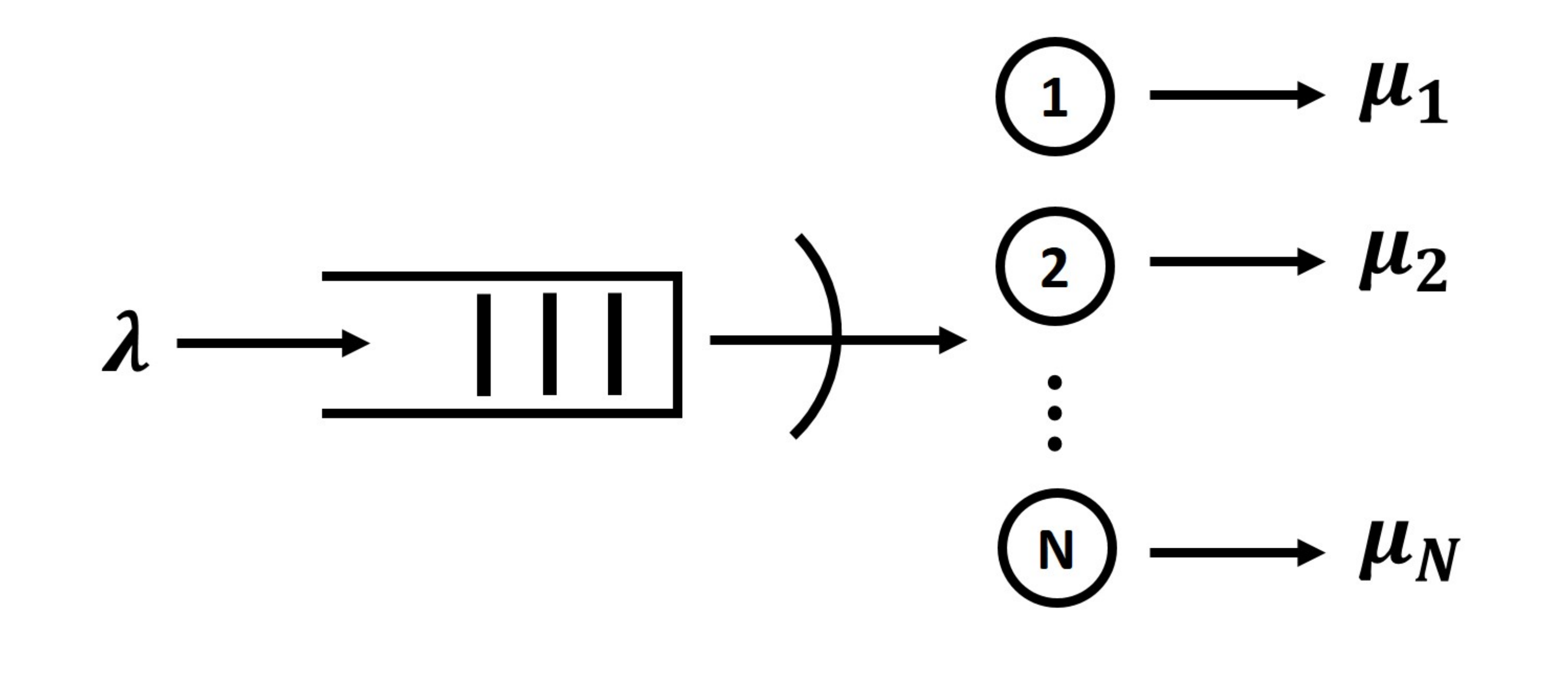}
\end{center}
  \caption{Queueing system with $N$ servers. At each time, one server may be scheduled. The rate at which each server can offer service is initially unknown.}
  \label{problem_setup_graphic}
\end{figure}

We consider a system consisting of a single queue and $N$ servers (where integer $N \in [2, \infty)$) operating over discrete time slots $t = 0, 1, 2, \dots$. See Fig.~\ref{problem_setup_graphic}. Packets arrive to the queue as a Bernoulli process $A(t)$ with rate $\lambda \in (0, 1]$ and wait in the queue until they receive service from one of the servers.  Each packet can be serviced no sooner than the next time slot after its arrival.  Each server's ability to offer service randomly fluctuates. Specifically, the number of packets that server $i \in [N]$ can service at time $t$ follows a Bernoulli process $D^i(t)$ with rate $\mu_i$. The arrival process and server processes are assumed to be independent of each other. We refer to server $i$ as stabilizing if $\mu_i > \lambda$, which implies that the rate at which the server can provide service keeps up with the arrival process.  Otherwise, we refer to it as non-stabilizing.  Given the above, an instantiation of our problem is characterized by the tuple $(\lambda, \vec{\mu})$, where $\vec{\mu}$ is the vector of service rates.  Then, we let $\mathcal{P}$ be the set of all problems (i.e., tuples).

The main challenge in our problem is that the system can only activate one server at a given time.
At each time slot, a system controller must select only one of the $N$ servers and ask it to offer service.  Then the selected server will inform the controller of the number of packets that it can offer to serve and, if the queue is non-empty, will service that number of packets.  We denote the controller's choice at time $t$ as $u(t) \in [N]$. For decision $u(t) = i$, the service offered to the queue is then denoted $D(t)$ which is equal to $D^i(t)$.  Throughout this work, the controller is not allowed to observe the offered service processes $D^i(t)$ prior to making its decision $u(t)$, and therefore it cannot know which servers will offer service prior to its choice.

Given the above, the queue backlog, $Q(t)$, evolves as
\begin{equation}
\label{how_the_queue_updates}
Q(t+1) = \left( Q(t) - D(t) \right)^+ + A(t), \mbox{ for } t = 0, 1, 2, \dots
\end{equation}
where $(x)^+$ is used to denote the maximum of $x$ and $0$. We assume $Q(0) = 0$ and the controller knows this.

The system incurs a unit cost, in delay, for each time slot that each packet has to wait in the queue.  Therefore, for a time horizon $T$, the controller wishes to minimize $\sum_{t = 0}^{T-1} Q(t)$.  Since, the controller cannot observe the values of $D^i(t)$ prior to making its decision $u(t)$,  the optimal action is to select the server
\begin{equation*}
i^* \triangleq \arg\max_{i \in [N]} \mu_i
\end{equation*}
to provide service to the queue. For simplicity, we assume $i^*$ is always unique.\footnote{Then, we let $\mathcal{P}$ be the set of all tuples such that $i^*$ is unique.}  In our framework, the controller does not a priori know the values of $\mu_i$ and must therefore use observations of $D(t)$ to identify $i^*$.  This implies that the service available from chosen server $u(t)$ is revealed to the controller after time $t$, but the service that would have been available from all other servers remains hidden.  Note that the controller can observe $D(t)$ at all times $t$, even when $Q(t) = 0$.  Finally, in this work, we assume that in addition to not knowing the values of $\mu_i$, the system does not initially know the exact value of $\lambda$, either.

Given the above, the objective is to design a controller that uses its observations to decide which server to schedule at each time slot to minimize 
\begin{equation*}
E\left[ \sum_{t=0}^{T-1} Q(t) \right].
\end{equation*}
To this end, denoting the history of all previous arrivals, service opportunities, and decisions as
\begin{equation*}
H(t)\!=\!( A(0), D(0), u(0), \dots, A(t-1), D(t-1), u(t-1) ),
\end{equation*}
we need to specify a controller policy $\pi$, which at each time $t$ uses any portion of $H(t)$ to (possibly randomly) make a decision on $u(t)$.  Note that since $H(t)$ contains the history of all previous arrivals and offered service to the queue, at time $t$, the policy implicitly knows the current backlog $Q(t)$ as well (i.e., we could explicitly include $Q(t)$ in $H(t)$, but this would be redundant).  We let $\Pi$ be the set of all policies that we could design.

Define $Q^*(t)$ to be the queue backlog under the controller that always schedules $i^*$, and let $Q^\pi(t)$ be the backlog under our policy $\pi$ that must learn which server is optimal.  We will analyze the performance of $\pi$ using the following definition of queue length regret,
\begin{equation}
\label{queue_regret}
R_{(\lambda, \vec{\mu})}^\pi(T) \triangleq E \left[ \sum_{t=0}^{T-1} Q^\pi(t) - \sum_{t=0}^{T-1} Q^*(t) \right],
\end{equation}
where the expectation is over the product measure of $P1 \times P2$ where $P1$ denotes the operation of $\pi$ and $P2$ the operation of the scheduler that always selects $i^*$. We will often simply refer to this metric as regret in the rest of the paper.  Observe that, if our policy minimizes \eqref{queue_regret} over the set $\Pi$, it must also minimize $E [\sum_{t=0}^{T-1} Q^\pi(t)]$ over the set.

\subsection{Analyzing Queue Length Regret}

Since at any time $t$, server $i^*$ has the highest probability of offering service to the queue, it is clear that the controller that always schedules $i^*$ must minimize the expected queue backlog. Therefore, for every time $t$ and any policy $\pi \in \Pi$, 
\begin{equation*}
E \left[Q^\pi(t)\right] \geq E \left[Q^*(t)\right],
\end{equation*}
which implies that $R_{(\lambda, \vec{\mu})}^\pi(T)$ is monotonically increasing in $T$ for all $\pi \in \Pi$.  Since, the queue begins empty at time $0$, using \eqref{how_the_queue_updates}, $R_{(\lambda, \vec{\mu})}^\pi(1) = R_{(\lambda, \vec{\mu})}^\pi(2) = 0$.

The focus in this work will be on characterizing how queue length regret can scale with time horizon $T$.  Therefore, we will not focus on finding policies that optimize $R_{(\lambda, \vec{\mu})}^\pi(T)$ for some finite value of $T$.  Instead, we will characterize $R_{(\lambda, \vec{\mu})}^\pi(T)$ as $T$ goes to infinity.  We will proceed to show that under different assumptions on $\lambda$ and $\vec{\mu}$, the regret $R_{(\lambda, \vec{\mu})}^\pi(T) = O(1)$.  Define the tail of the regret to be 
\begin{equation*}
L^\pi_{(\lambda, \vec{\mu})}(T) \triangleq \sum_{t=T}^{\infty} E \left[  Q^\pi(t) - Q^*(t) \right].
\end{equation*}
Intuitively, this is the additional cost-to-go of infinitely running the policy past time $T$. Then the following proposition holds.
\begin{proposition}
If for policy $\pi \in \Pi$ and subset $\mathcal{P}' \subseteq \mathcal{P}$, $R_{(\lambda, \vec{\mu})}^\pi(T) = O(1)$ for each $(\lambda, \vec{\mu}) \in \mathcal{P}'$, then $L^\pi_{(\lambda, \vec{\mu})}(T) \to 0$ as $T \to \infty$ for every $(\lambda, \vec{\mu}) \in \mathcal{P}'$.
\end{proposition} 
Thus, we see that a policy that obtains $O(1)$ queue length regret seeks to minimize the tail of the regret as time goes to infinity (i.e., causes the tail to trend to zero). A policy that optimizes the tail of the regret is not guaranteed to minimize the queue length regret. Indeed the policies we proceed to analyze in Section~\ref{section_1} do not generally achieve small queue length regret, and are instead designed to facilitate proving that $O(1)$ queue length regret is achievable.  Nevertheless, we will show in Section~\ref{section_simulation} that the insight gained from these policies can be used to construct heuristics with good regret performance.

\section{Regret of Traditional Bandit Algorithms}
\label{section_regret_bandit_alg}
In this section, we establish that the regret of any policy $\pi \in \Pi$ that does not use previous observations of the arrival process must have a queue length regret that grows logarithmically with $T$ for some $\vec{\mu}$.  Under this restriction, the policies in this section (in effect) do not observe previous arrivals to the queue, and their decision making process is solely focused on the observed offered services. Since the policies do not monitor process $A(t)$, they cannot directly use the queue backlog $Q(t)$ to make their decisions, either.  Under this restriction, we may still borrow any strategy from the traditional stochastic multi-armed bandit literature to solve the problem.  These policies focus on only maximizing offered service to the queue, using previous observations of the offered service to guide their decisions.  Without loss of generality, the theorem is established for a system with two servers.  The theorem's proof makes use of a well-known lower bound on the performance of bandit learning algorithms \cite[Theorem 2.2]{bubeck}.

Let $\mu_{(j)}$ be the $j^{th}$ service rate when the service rates are sorted in increasing order.

\begin{theorem}
\label{theorem0}
For any  $\pi \in \Pi$ that does not use previous observations of the arrival process, $\exists \vec{\mu}$ with server rates $\mu_{(1)}$ and $\mu_{(2)}$ ($\mu_{(2)} > \mu_{(1)}$) such that for any fixed $\lambda \in \left(0, \mu_{(2)} \right)$, $R_{(\lambda, \vec{\mu})}^\pi(T) = \Omega(\log{T})$.
\end{theorem}

Note that to facilitate later comparison to Theorem~\ref{theorem3}, the theorem statement specifies $\lambda \in (0, \mu_{(2)})$, but the proof actually holds for the larger set $\lambda \in (0, 1]$.  We let $\mathbf{1}\{\cdot\}$ be the indicator random variable.

\begin{IEEEproof}
We apply a sample path argument.  There are two servers in the system. Denote the optimal server as $(2)$ and the suboptimal server as $(1)$.   Consider an arbitrary policy $\pi$ that does not make its decisions using observations of the arrival process.  Define $\tilde{\pi}(t)$ to be the decision process of a controller that follows $\pi$ up to time $t-1$, but at time $t$ chooses server $(2)$ with probability $1$.  Let $Q^\pi(\tau)$ and $Q^{\tilde{\pi}(t)}(\tau)$ be the queue backlogs under the two respective controllers.

Then, for any outcome $\omega$ defining a fixed sample path for the arrival process, service processes, and possible randomization in $\pi$
\begin{equation}
\label{first_equation_of_theorem0}
Q^\pi(\tau, \omega) = Q^{\tilde{\pi}(t)}(\tau, \omega), \forall \tau \leq t.
\end{equation}
Denote the offered service from servers $(1)$ and $(2)$ as $D^{(1)}(t, \omega)$ and $D^{(2)}(t, \omega)$ respectively and the decision by policy $\pi$ at time $t$ as $u(t, \omega)$.  Then, from \eqref{first_equation_of_theorem0} and the queue evolution equation,
\begin{equation*}
Q^\pi(t + 1, \omega) - Q^{\tilde{\pi}(t)}(t + 1, \omega) = \left\{
\begin{array}{ll}
D^{(2)}(t, \omega) - D^{(1)}(t, \omega), &\mbox{ if } \left\{Q^\pi(t, \omega) > 0\right\} \cap \left\{u(t, \omega) = (1)\right\} \\
0, & \mbox{ otherwise}
\end{array}
\right. .
\end{equation*}

Thus, taking expectation and summing over time
\begin{multline}
\label{bounding_expectation1}
E\left[ \sum_{t = 2}^{T - 1} Q^\pi(t) - \sum_{t = 2}^{T - 1} Q^{\tilde{\pi}(t-1)}(t) \right] =
E\left[ \sum_{t = 1}^{T - 2} Q^\pi(t + 1) - \sum_{t = 1}^{T - 2} Q^{\tilde{\pi}(t)}(t + 1) \right] \\ = E \left[ \sum_{t = 1}^{T-2} \left( D^{(2)}(t) - D^{(1)}(t) \right) \mathbf{1}\left\{ \left\{Q^\pi(t) > 0\right\} \cap \left\{u(t) = (1)\right\} \right\} \right].
\end{multline}
Note that each $\tilde{\pi}(t)$ defines a different decision process leading up to time $t$.  For example, $\tilde{\pi}(t-1)$ chooses server $(2)$ with probability $1$ at time $t-1$, but $\tilde{\pi}(t)$ follows the actions of policy $\pi$ at time $t-1$ (and chooses server $(2)$ with probability $1$ at time $t$, instead).  One can then see that in the above, at each index $t$, we are comparing the performance of $\pi$ to the performance of a different $\tilde{\pi}(t)$ decision process. Since the queue begins empty at $t=0$, by \eqref{how_the_queue_updates}, the expected queue backlog under any controller must be equal to $0$ and $\lambda$ at times $0$ and $1$, respectively.  Thus, the left-hand side of \eqref{bounding_expectation1} has been chosen as a summation from time slot $2$ to $T-1$.

Now, $Q^\pi(t)$ and $u(t)$ are functions of events up to time $t-1$ and are therefore independent of $D^{(1)}(t)$ and $D^{(2)}(t)$.  Thus, we can write
\begin{equation}
\label{bounding_expectation2}
\eqref{bounding_expectation1} = (\mu_{(2)} - \mu_{(1)}) \sum_{t = 1}^{T - 2} P\left( \left\{Q^\pi(t) > 0\right\} \cap \left\{u(t) = (1)\right\} \right).
\end{equation}
Now, $\left\{Q^\pi(t) > 0\right\}$ and $\left\{u(t) = (1)\right\}$ are not necessarily independent events.  However, we can use the fact that an arrival at time $t-1$ is a subset of the event that $Q^\pi(t) > 0$ to lower bound the above (i.e., $\left\{A(t - 1) = 1\right\} \subseteq \left\{ Q^\pi(t) > 0 \right\}$).  Then,
\begin{equation}
\label{bounding_expectation3}
\eqref{bounding_expectation2} \geq \left(\mu_{(2)} - \mu_{(1)}\right) \sum_{t = 1}^{T-2} P\left( \left\{A(t-1) = 1\right\} \cap \left\{u(t) = (1) \right\} \right).
\end{equation}
We then use the fact that policy $\pi$ is independent of the arrival process $A(t)$ to obtain
\begin{equation}
\label{bounding_expectation4}
\eqref{bounding_expectation3} = \lambda \left(\mu_{(2)} - \mu_{(1)}\right) \sum_{t = 1}^{T-2} P\left( u(t) = (1) \right).
\end{equation}
Note that $\lambda \left(\mu_{(2)} - \mu_{(1)}\right) > 0$.
Now, policy $\pi$ is a strategy for solving the stochastic multi-armed bandit problem (i.e., it uses the previous actions and observed offered service to decide on which server to schedule at time $t$).  From \cite[Theorem 2.2]{bubeck}, there must exist a $\vec{\mu}$ such that\footnote{This is because either: the policy has $\sum_{t = 0}^{T-1} P\left( u(t) = (1) \right) = \Omega\left(T^a\right)$ for some $a > 0$ and some Bernoulli distribution on the rewards (which we can assume to be $\vec{\mu}$, without loss of generality), or $\sum_{t = 0}^{T-1} P\left( u(t) = (1) \right) = \Omega(\log{T})$ for any Bernoulli distribution on the rewards. See \cite[Theorem 2.2]{bubeck} for details.}
\begin{equation*}
\sum_{t = 0}^{T-1} P\left( u(t) = (1) \right) = \Omega(\log{T}).
\end{equation*}
 Without loss of generality, we can then assume the chosen $\vec{\mu}$ has this property. Using equations \eqref{bounding_expectation1} through \eqref{bounding_expectation4}, this implies
\begin{equation}
\label{bounding_expectation5}
E\left[ \sum_{t = 1}^{T - 2} Q^\pi(t + 1) - \sum_{t = 1}^{T - 2} Q^{\tilde{\pi}(t)}(t + 1) \right] = \Omega(\log{T}).
\end{equation}
Then, because $E\left[Q^*(t + 1)\right] \leq E\left[Q^{\tilde{\pi}(t)}(t + 1)\right]$ for all $t$,\footnote{It is easy to show through dynamic programming that, when $\vec{\mu}$ is given to the controller a priori, the controller that minimizes the expected queue backlog at each time $t$, schedules $i^*$ for all time $t$.}
\begin{equation*}
E\left[ \sum_{t = 1}^{T - 2} Q^\pi(t + 1) - \sum_{t = 1}^{T - 2} Q^*(t + 1) \right] \geq E\left[ \sum_{t = 1}^{T - 2} Q^\pi(t + 1) - \sum_{t = 1}^{T - 2} Q^{\tilde{\pi}(t)}(t + 1) \right].
\end{equation*}
Therefore, by \eqref{bounding_expectation5}, 
\begin{equation*}
E\left[ \sum_{t = 1}^{T - 2} Q^\pi(t + 1) - \sum_{t = 1}^{T - 2} Q^*(t + 1) \right] = \Omega(\log{T}).
\end{equation*}
Using the definition of $R_{(\lambda, \vec{\mu})}^\pi(T)$ given by \eqref{queue_regret}, this proves the result.
\end{IEEEproof}

\section{Regret of Queue-Length-Based Policies}
\label{section_1}

In this section, we examine the asymptotic scaling of regret $R_{(\lambda, \vec{\mu})}^\pi(T)$ for policies that make decisions using $Q(t)$ (or equivalently, the previous observations of the arrival process).  We do so by considering the problem for different subsets of $\mathcal{P}$ that impose different assumptions on the relationship between $\lambda$ and $\vec{\mu}$.  This section will build to the main result of this work, Theorem~\ref{theorem3}, which states that there exists a $\pi \in \Pi$ such that for any problem $(\lambda, \vec{\mu})$ with $\mu_{i^*} > \lambda$, $R_{(\lambda, \vec{\mu})}^\pi(T) = O(1)$.

We begin our analysis in Subsection~\ref{section_1_1} under the assumption that every server is stabilizing (i.e., $\mu_i > \lambda$, $\forall i \in [N]$). Under this assumption, the controller does not need to account for the possibility of system instability.  As a result, the controller can limit itself to performing exploration only on time slots in which the queue is empty.  During time slots in which the queue is backlogged, the controller will exploit its previously obtained knowledge to schedule the server it believes to be best.

In Subsection~\ref{section_1_2}, we allow for both stabilizing and non-stabilizing servers in the system (i.e., $\mu_i$ is allowed to be less than $\lambda$ for a non-empty subset of the servers).  However, we will assume that the controller is given a randomized policy that achieves an offered service rate that is greater than the arrival rate to the queue.  Note that the controller does not need to learn this policy and can use it to return the queue to the empty state.  The given randomized policy is not required to have any relationship to $i^*$ and will not, in general, minimize queue length regret.  As a result, to minimize queue length regret, the controller will not want to excessively rely upon it.

In Subsection~\ref{section_1_3}, we further relax our assumptions on the problem and only require that $\mu_{i^*} > \lambda$.  In this subsection, the controller will need to identify which servers are stabilizing, while simultaneously trying to minimize $R_{(\lambda, \vec{\mu})}^\pi(T)$.  This will require a policy that does not destabilize the system.  To this end, the controller will have to explore the servers' offered service rates during both time slots when the queue is empty and backlogged.  Explorations during time slots when the queue is backlogged, in general, waste work and should therefore be performed sparingly.  Intuitively, as the controller identifies which subset of servers have service rates $\mu_i > \lambda$, it can focus its explorations on time slots when the queue is empty.

The above three cases build upon one another.  The insight from one will point to a policy for the next, and we will therefore analyze the above cases in sequence.  Under each of the above assumptions, we will find that there exists a policy such that the regret converges for all $(\lambda, \vec{\mu})$ meeting the assumption.  In contrast, in Section~\ref{section_1_4}, we show that there does not exist a policy that can achieve convergent regret over the class of all problems for which $\mu_i \leq \lambda$, $\forall i \in [N]$.

Note that the main difficulty in evaluating a policy's performance is the correlation between the time-evolving queue backlog $Q(t)$ and the history of observed offered service, which is used to find the best server.  In general, the queue will be empty more often once we begin to identify the best server, and, likewise, the best server will be more easily identified when we begin to obtain more empty periods in which we can freely explore.  The policies considered in this section have been chosen to disentangle this relationship so that well-known tools from statistics can be applied in our analysis. Specifically, our results will make use of Hoeffding's inequality, which we restate here for convenience.

Consider $J_n = I^1 + I^2 + \dots + I^n$ where $I^k$ are i.i.d. random variables with mean $\overline{I}$ and support between $[x_1, x_2]$ for $x_1 < x_2$.  Then, by Hoeffding's inequality, for all $x > 0$,
\begin{align*}
P\left(\frac{1}{n}J_n - \overline{I} \geq x \right) &\leq e^{-2n \frac{x^2}{(x_2 - x_1)^2}} \\ \nonumber
P\left(\frac{1}{n}J_n - \overline{I} \leq -x \right) &\leq e^{-2n \frac{x^2}{(x_2 - x_1)^2}}. \nonumber
\end{align*}
For simplicity, we will often use Hoeffding's inequality in our analysis even when tighter concentration inequalities may exist.

\subsection{All Servers Are Stabilizing}
\label{section_1_1}

In this subsection, we assume all servers are stabilizing and design policies that are able to obtain $O(1)$ regret under this assumption.  Specifically, we will assume that the environment will only choose parameters $(\lambda, \vec{\mu})$ from a subset $\mathcal{P}_1$ defined below.
\begin{assumption}
\label{assumption1}
$\left(\lambda, \vec{\mu} \right) \in \mathcal{P}_1 \triangleq \left\{ \mathcal{P} : \mu_i > \lambda, \forall i \in [N] \right\}$.
\end{assumption}
Under this assumption we will prove the following theorem, which states that there exists a policy such that, for every $(\lambda, \vec{\mu})$ in the set $\mathcal{P}_1$, the regret converges.  The proof of Theorem~\ref{theorem1} is constructive and will give a policy for obtaining the result.
\begin{theorem}
\label{theorem1}
Under Assumption~\ref{assumption1}, $\exists \pi \in \Pi$ such that, for each $(\lambda, \vec{\mu}) \in \mathcal{P}_1$, $R_{(\lambda, \vec{\mu})}^\pi(T) = O(1)$.
\end{theorem}

\begin{figure}
\noindent\makebox[\linewidth]{\rule{\columnwidth}{0.4pt}}
\begin{algorithmic}
\STATE{ $\widehat{\mu}_i = 0$ for all $i \in [N]$}
\WHILE{TRUE}
	\FOR{Empty period $p$}
		\STATE{At the first time slot of the period, set $u(t) = i$ uniformly at random over $[N]$ and update $\widehat{\mu}_i$ with the observed state of $D(t)$}
		\STATE{For other time slots in the empty period, idle}
	\ENDFOR
	\FOR{Busy period $p$}
		\STATE{Schedule $\arg \max_{i \in [N]} \widehat{\mu}_i$ until the queue empties }
	\ENDFOR
\ENDWHILE
\end{algorithmic}
\noindent\makebox[\linewidth]{\rule{\columnwidth}{0.4pt}}
\caption{Policy $\pi_1$ for achieving Theorem~\ref{theorem1}.  Each variable $\widehat{\mu}_i$ is initialized to zero prior to its first update.}
\label{policy1}
\end{figure}

To prove Theorem~\ref{theorem1}, we analyze the policy $\pi_1$ shown in Fig.~\ref{policy1} on an arbitrary $(\lambda, \vec{\mu}) \in \mathcal{P}_1$.  This policy maintains sample mean variables $\widehat{\mu}_i$ that estimate the servers' service rates $\mu_i$.  Each sample mean $\widehat{\mu}_i$ is updated using observations of server $i$'s offered service. However, it is important to note that each sample mean is not updated at every time slot that we schedule its corresponding server, and we will therefore not use every observation of the offered services to construct the sample means.  Instead, to facilitate analysis, we will only update the sample means at strategically chosen time slots.  Throughout this section, sample mean variables are initialized to zero prior to their first update with a service observation.  After the first update, the sample mean variables equal the sum of the used observations divided by the number of used observations.

\begin{figure}
\begin{center}
    \includegraphics[width=0.9\linewidth]{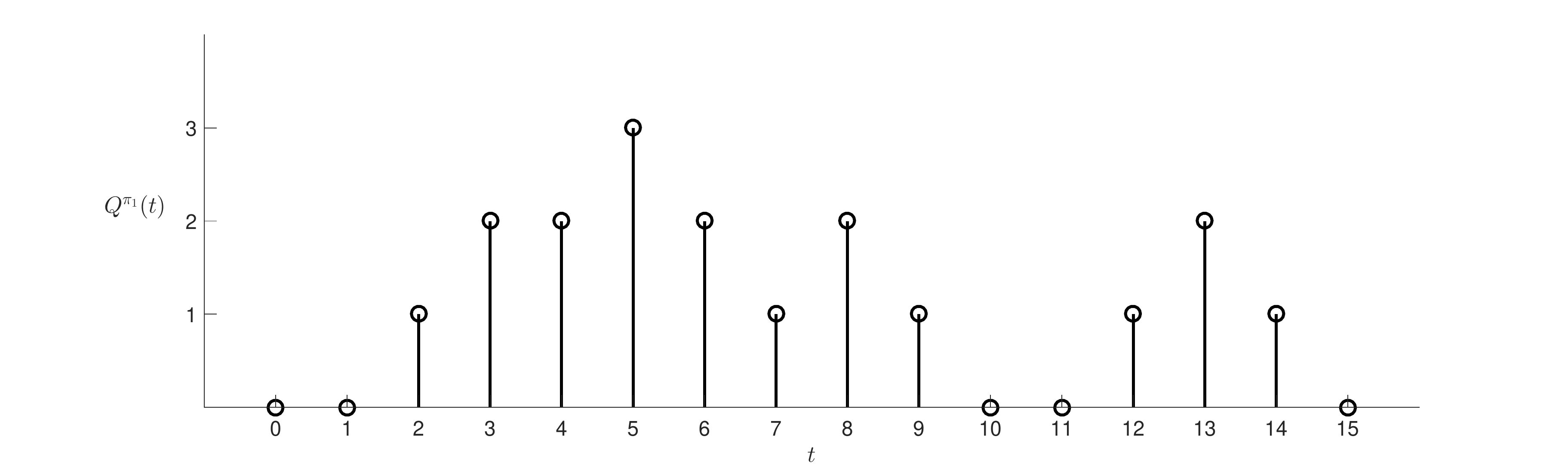}
\end{center}
  \caption{Example sample path for $Q^{\pi_1}(t)$ over the first $16$ time slots for some, unspecified, $(\lambda, \vec{\mu})$.  Empty period $1$ occurs over time slots $0$ and $1$. Busy period $1$ begins in time slot $2$ and ends with time slot $9$.  The duration of the busy period is $8$ time slots, and the integral of the queue backlog over the busy period is $14$.  Empty period $2$ begins at time slot $10$, and busy period $2$ begins at time slot $12$. The integral of the queue backlog over busy period $2$ is $4$.  Finally, empty period $3$ is shown beginning in time slot $15$.}
  \label{queue_evolution_graphic}
\end{figure}

Note that under policy $\pi_1$, the queue backlog will transition through alternating time intervals, wherein the queue is continuously empty over an interval of time (i.e., $Q^{\pi_1}(t) = 0$) and then continuously busy over an interval of time (i.e., $Q^{\pi_1}(t) > 0$).  We enumerate the periods using positive integers $p = 1, 2, \dots$ and refer to the $p^{th}$ occurrence of an empty period (busy period) as empty period $p$ (busy period $p$, respectively).  To arrive at Theorem~\ref{theorem1}, we will analyze the integrated queue backlog taken over the busy periods.  To this end, we define the \emph{integral of the queue backlog over busy period} $p$ to be the summation of the queue backlog $\sum_t Q^{\pi_1}(t)$ taken over those times $t$ that are in busy period $p$.  By our problem formulation, a unit cost is incurred by the system for each time slot each packet waits in the queue, and the integral of the queue backlog over busy period $p$ can then be interpreted as the total cost accumulated over the busy period.  In Fig.~\ref{queue_evolution_graphic}, we illustrate our terminology on a sample path of $Q^{\pi_1}(t)$.

\begin{figure}
\begin{center}
    \includegraphics[width=0.9\linewidth]{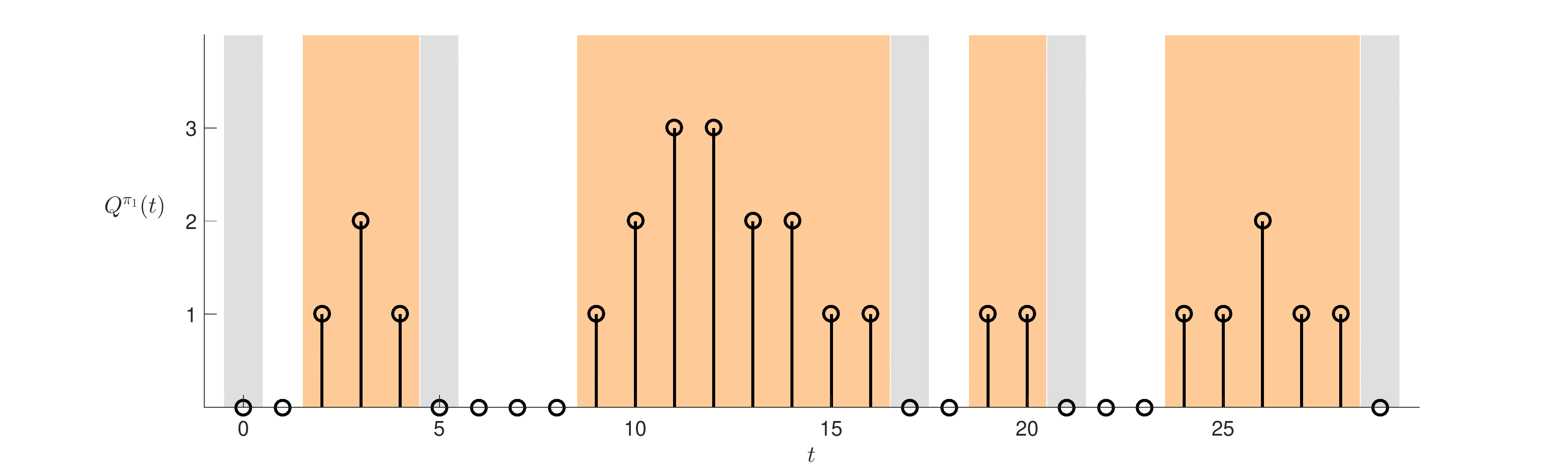}
\end{center}
  \caption{Example sample path of $Q^{\pi_1}(t)$ evolving over the first $30$ time slots for some, unspecified, $(\lambda, \vec{\mu})$. The background colors indicate the type of action policy $\pi_1$ took during the corresponding time slots (see Fig.~\ref{policy1}).  During time slots marked in gray, $\pi_1$ randomly selected one of the servers $i$ and updated its sample mean $\widehat{\mu}_i$ with the observed offered service.  The sample means were only updated during these time slots.  During time slots marked in orange, $\pi_1$ scheduled the server with the (so-far) largest sample mean to service the queue.  During time slots with a white background, no action was taken.}
  \label{pi_1_illustrated_figure}
\end{figure}

We now briefly point out some aspects of policy $\pi_1$.  See Fig.~\ref{pi_1_illustrated_figure}, which provides an illustration of $\pi_1$.  Under the policy, the first time slot of each empty period is used to update one of the sample means.  To do this, the policy chooses one of the servers uniformly at random and schedules it.  The amount of service that is offered from the chosen server is then observed and used to update the server's corresponding sample mean.  Note that the sample means are only updated during these time slots, and all other observations of the servers' offered services, that are made during other time slots, are not used in our estimations. Now, following an empty period, when a packet arrives to the empty queue, the system enters a busy period.  At the start of each busy period, the policy chooses to schedule the server with the highest sample mean (breaking ties between sample means according to any arbitrary decision rule) and continues to schedule that server until the queue finally empties and the busy period ends.  Thus, during each busy period the queue is serviced by only one server (namely, the one with highest sample mean).  We again stress that the policy does not use the observations of the service obtained during the busy period to update the sample mean.  Notionally, the policy performs no exploration during busy periods and instead focuses on exploiting previous observations to empty the queue quickly.

Now, suppose we are at the beginning of busy period $p$ and policy $\pi_1$ chooses to schedule server $i$ for this busy period.  Then, the duration of the busy period (i.e., the amount of time that will be required to empty the queue) is given by a random variable $X_i$, whose mean we denote $\overline{X}_i$.  Note that conditioned on server $i$ being scheduled during the busy period, $X_i$ is not a function of the busy period number $p$.  By Assumption~\ref{assumption1}, for all $i \in [N]$ we have that $\mu_i > \lambda$ and therefore $\overline{X}_i$ is finite.  Now, to establish the proof of Theorem~\ref{theorem1}, we will be interested in the integral of the queue backlog taken over the busy period.  Note that, as with the busy period's duration, the integral of the queue backlog is given by a random variable, which we denote $Z_i$. Using $\tau$ to denote the arbitrary start of the busy period,
\begin{equation*}
Z_i \triangleq \sum_{t = \tau}^{\tau + X_i - 1} Q^{\pi_1}(t)
\end{equation*}
and has mean $\overline{Z}_i$.
As with the busy period duration, variable $Z_i$ is not a function of the busy period number $p$.

We introduce one last piece of notation before moving to the proof of Theorem~\ref{theorem1}.  At the start of each busy period, policy $\pi_1$ schedules the server with the highest sample mean to empty the queue (i.e., the server with the highest sample mean is scheduled to every time slot in the busy period until the queue empties).  We will refer to this decision as $\pi_1$ scheduling the busy period to server $i$.  In the following analysis, we will be interested in the frequency with which the policy schedules each server $i$ to busy periods. Therefore, we will use $S^{\pi_1}_i(P)$ to denote the number of busy periods that $\pi_1$ schedules to server $i$ in the first $P$ busy periods.  We will use these variables to bound, in expectation, the number of busy periods that are scheduled to a server other than $i^*$.

The proof of Theorem~\ref{theorem1} now proceeds through four lemmas.  We begin with Lemma~\ref{lemma1}.  We denote the integral of the queue backlog taken over busy periods $\pi_1$ schedules to $i$ as:
\begin{equation}
\sum\limits_{t = 0}^{T - 1} Q^{\pi_1}(t) \mathbf{1} \left\{ u(t) = i \right\}.
\label{Q_indicator}
\end{equation}
For example, if in Fig.~\ref{pi_1_illustrated_figure} the first and fourth busy periods were scheduled to server $1$ but the second and third busy periods were scheduled to server $2$, then $\sum_{t = 0}^{29} Q^{\pi_1}(t) \mathbf{1} \left\{ u(t) = 1 \right\} = 10$ and $\sum_{t = 0}^{29} Q^{\pi_1}(t) \mathbf{1} \left\{ u(t) = 2 \right\} = 17$.   Lemma~\ref{lemma1} shows that we can upper bound the queue length regret with the queue backlog summed over times that $\pi_1$ is not scheduling $i^*$.  The proof follows from a sample path argument.

Consider a fixed outcome  $\omega$ for the arrival and service processes and randomization in policy $\pi_1$.  Then, $Q^{\pi_1}(t, \omega)$ is the queue backlog of policy $\pi_1$ at time $t$ for this sample path, and $Q^*(t, \omega)$ is the backlog at time $t$ for a controller that schedules $i^*$ for all time slots since the start of time for this sample path. Now, assume at time $t$ policy $\pi_1$ is in a busy period where it is scheduling $i^*$.  Then, $Q^{\pi_1}(t,\omega) \leq Q^*(t, \omega)$.  The intuition behind this claim follows from the fact that the queue is empty right before the busy period begins and $\pi_1$ does not switch servers mid-busy period.  Therefore, for the given sample path of arrivals and service processes, had the controller scheduled $i^*$ for all time slots since the start of time, the queue could not be less.  Taking expectations and using the definition of queue length regret in \eqref{queue_regret} then gives the result.  The complete proof of the lemma can be found in Appendix~\ref{lemma1_appendix}.
\begin{lemma}
\label{lemma1}
\begin{equation*}
R_{(\lambda, \vec{\mu})}^{\pi_1}(T) \leq E \left[ \sum\limits_{t = 0}^{T - 1} \sum\limits_{i \in [N] - i^*} Q^{\pi_1}(t) \mathbf{1} \left\{ u(t) = i \right\} \right].
\end{equation*}
\end{lemma}

Next, in Lemma~\ref{lemma2}, the expected value of $Z_i$ is shown to be finite.  Note that by our problem's definition at most one packet can arrive to the queue at each time slot.  Then, since a busy period must start with one packet in queue at its first time slot (i.e., the packet that arrives to the queue to begin the busy period), the maximum size of the queue backlog during the busy period cannot be larger than the busy period's duration.   Thus, the expected value of $Z_i$ can be bounded by the expected value of $X_i^2$, which is finite.  The complete proof of the lemma can be found in Appendix~\ref{lemma2_appendix}.

\begin{lemma}
\label{lemma2}
$\overline{Z}_i < \infty$.
\end{lemma}

We next proceed to bound the expected value of \eqref{Q_indicator} with Lemma~\ref{lemma3}.  The lemma states that the integrated queue backlog (taken over busy periods scheduled to server $i$) up until time $T$ is less than the integrated queue backlog (taken over busy periods scheduled to server $i$) through the $T^{th}$ busy period.    Observe that the right hand side of \eqref{Q_indicator_bound}, is the expected number of busy periods out of $T$ busy periods scheduled to server $i$, $E\left[ S^{\pi_1}_i(T) \right]$, multiplied by the expected integrated queue backlog, $\overline{Z}_i$. This quantity is the expected value of the integral of the queue backlog, over busy periods scheduled to $i$, up to busy period $T$.

The intuition behind Lemma~\ref{lemma3} is simple.  Since, by definition, each busy period must be at least one time slot in duration, we cannot have more than $T$ busy periods in $T$ time slots.  (In fact, since empty periods must also be one time slot in duration, it is obvious that we will have far fewer than $T$ busy periods in any $T$ time slots.)  Therefore, for any outcome of the system's performance, the integrated queue backlog up to time slot $T$ will be less than the integrated queue backlog through the $T^{th}$ busy period.  Taking expectations then gives the result.  The complete proof of the lemma can be found in Appendix~\ref{lemma3_appendix}.
\begin{lemma}
\label{lemma3}
\begin{equation}
\label{Q_indicator_bound}
E \left[ \sum_{t = 0}^{T - 1} Q^{\pi_1}(t) \mathbf{1} \left\{ u(t) = i \right\} \right] \leq \overline{Z}_i E \left[ S^{\pi_1}_i(T) \right].
\end{equation}
\end{lemma}

Finally, in Lemma~\ref{lemma4}, we show that the expected number of times $\pi_1$ schedules server $i \neq i^*$ to busy periods is bounded by a finite constant which is independent of $T$.  In order to schedule server $i \neq i^*$ to a busy period, $\pi_1$ must estimate $\widehat{\mu}_i$ to be no less than $\widehat{\mu}_{i^*}$.  Since we obtain a new observation of one of the servers during each empty period $p$, using Hoeffding's inequality, we can show that the probability that there exists a server $i \neq i^*$ such that $\widehat{\mu}_i \geq \widehat{\mu}_{i^*}$ decays exponentially in $p$.  The result then follows from the convergence of geometric series.  The proof of the lemma can be found in Appendix~\ref{lemma4_appendix}.
\begin{lemma}
\label{lemma4}
$E\left[ S^{\pi_1}_i(T) \right] = O(1)$, $\forall i \in [N] - i^*$.
\end{lemma}

Given the above four lemmas, we are now ready to establish Theorem~\ref{theorem1}.

\begin{IEEEproof}[Proof of Theorem~\ref{theorem1}]
Combining Lemmas~\ref{lemma1}~and~\ref{lemma3}, 
\begin{equation*}
R_{(\lambda, \vec{\mu})}^{\pi_1}(T) \leq \sum_{i \in [N] - i^*} \overline{Z}_i E \left[ S^{\pi_1}_i(T) \right].
\end{equation*}
  Then by Lemmas~\ref{lemma2}~and~\ref{lemma4}, $R_{(\lambda, \vec{\mu})}^{\pi_1}(T) = O(1)$ giving the result.
\end{IEEEproof}

As a final note, it is important to observe that, by its statement, Theorem~\ref{theorem1} does not imply that there exists a constant that bounds $R_{(\lambda, \vec{\mu})}^{\pi_1}(T)$ for all $(\lambda, \vec{\mu}) \in \mathcal{P}_1$.  It only states that for any chosen $(\lambda, \vec{\mu}) \in \mathcal{P}_1$, under policy $\pi_1$, the queue length regret converges to a finite number that is dependent on the chosen parameters.

\subsection{Non-Stabilizing Servers}
\label{section_1_2}

In this subsection, we relax Assumption~\ref{assumption1} to allow for non-stabilizing servers.  Thus, we will now allow $\mu_i \leq \lambda$ for some strict subset of the servers.  Importantly, we will assume that the controller is given a known convex summation over the servers' rates that strictly dominates the arrival rate to the queue.  Then, by randomizing over the servers using this convex summation, the controller can always stabilize the system.  Concretely, we make the following assumption.  

\begin{assumption}
\label{assumption2}
For some given $\alpha_i \geq 0$ and $\sum_{i \in [N]} \alpha_i = 1$,
\begin{equation}
(\lambda, \vec{\mu}) \in \mathcal{P}_2(\vec{\alpha}) \triangleq \Big\{ \mathcal{P} : \sum_{i \in [N]} \alpha_i \mu_i > \lambda\Big\}.
\label{convex_sum_over_mu}
\end{equation}
\end{assumption}
Observe that the set $\mathcal{P}_2(\vec{\alpha})$ defined above is determined by the values of $\alpha_i$ that are given to the controller and different values will lead to different sets.  We emphasize that Assumption~\ref{assumption2} does not imply that the controller knows any of the values of $\mu_i$, a priori.  Rather, it states that the controller can be confident that the convex summation it is provided will meet the condition of \eqref{convex_sum_over_mu}.

For example, if $\alpha_1 = 1$ then Assumption~\ref{assumption2} reduces to $\mu_1 > \lambda$ and this fact being known by the controller, a priori.  Then, if at any given moment the controller wishes to return the queue to the empty state, it can simply resort to persistently scheduling server $1$ and wait for the queue to empty.  However, note that server $1$ may not be the optimal server $i^*$, and, as a result, persistently scheduling server $1$ for every time slot will not in general give good queue length regret performance for every $(\lambda, \vec{\mu}) \in \mathcal{P}_2(\vec{\alpha})$.

Generalizing the above example to arbitrary $\alpha_i$ values, we see that Assumption~\ref{assumption2} implies that, if at any point the controller decides to resort to a strategy of scheduling at each time slot server $i$ with probability $\alpha_i$, it can be guaranteed that the queue will eventually empty.  In other words, the values of $\alpha_i$ define a stationary, randomized policy that has a service rate that dominates the arrival rate to the queue and can, therefore, empty the queue infinitely often.  However, since by the assumption's statement the $\alpha_i$ values are not required to have any special relationship to $i^*$, the randomized policy will not generally minimize regret, and its use should not be overly relied upon.

Before continuing, we note that Assumption~\ref{assumption2} is intentionally artificial.  The assumption assumes a randomized policy is given to the controller and that the system will only encounter parameters for $\lambda$ and $\vec{\mu}$ such that the given policy can stabilize the queue.  In most practical applications, this assumption is unrealistic.  However, the results of this subsection will be highly instructive for the next subsection, where we relax Assumption~\ref{assumption2} and no longer assume a known stabilizing policy is given to the controller.  Importantly, in the following, we will develop a method that can sparingly use a suboptimal policy to consistently guarantee future empty periods for the queue without overly using the suboptimal policy so as to lose the bounded regret result.    

We now give the main theorem of this subsection.

\begin{figure}
\noindent\makebox[\linewidth]{\rule{\columnwidth}{0.4pt}}
\begin{algorithmic}
\STATE{$\widehat{\mu}_i = 0$ for all $i \in [N]$}
\WHILE{TRUE}
	\FOR{Empty period $p$}
		\STATE{At the first time slot of the period, set $u(t) = i$ uniformly at random over $[N]$ and update $\widehat{\mu}_i$ with the observed state of $D(t)$}
		\STATE{For other time slots in the empty period, idle}
	\ENDFOR
	\FOR{Busy period $p$}
		\STATE{Schedule $\arg \max_{i \in [N]} \widehat{\mu}_i$ for the first $p$ time slots of the busy period or until the queue empties}
		\IF{The queue does not empty during the first $p$ time slots}
			\STATE{At each time slot, schedule server $i$ with probability $\alpha_i$ until the queue empties}
		\ENDIF
	\ENDFOR
\ENDWHILE
\end{algorithmic}
\noindent\makebox[\linewidth]{\rule{\columnwidth}{0.4pt}}
\caption{The policy $\pi_2$ for achieving Theorem~\ref{theorem2}.  Each variable $\widehat{\mu}_i$ is initialized to zero prior to its first update.}
\label{policy_assumption_2}
\end{figure}

\begin{theorem}
\label{theorem2}
Under Assumption~\ref{assumption2}, $\exists \pi \in \Pi$ such that, for each $(\lambda, \vec{\mu}) \in \mathcal{P}_2(\vec{\alpha})$, $R_{(\lambda, \vec{\mu})}^\pi(T) = O(1)$.
\end{theorem}

\begin{figure}
\begin{center}
    \includegraphics[width=0.9\linewidth]{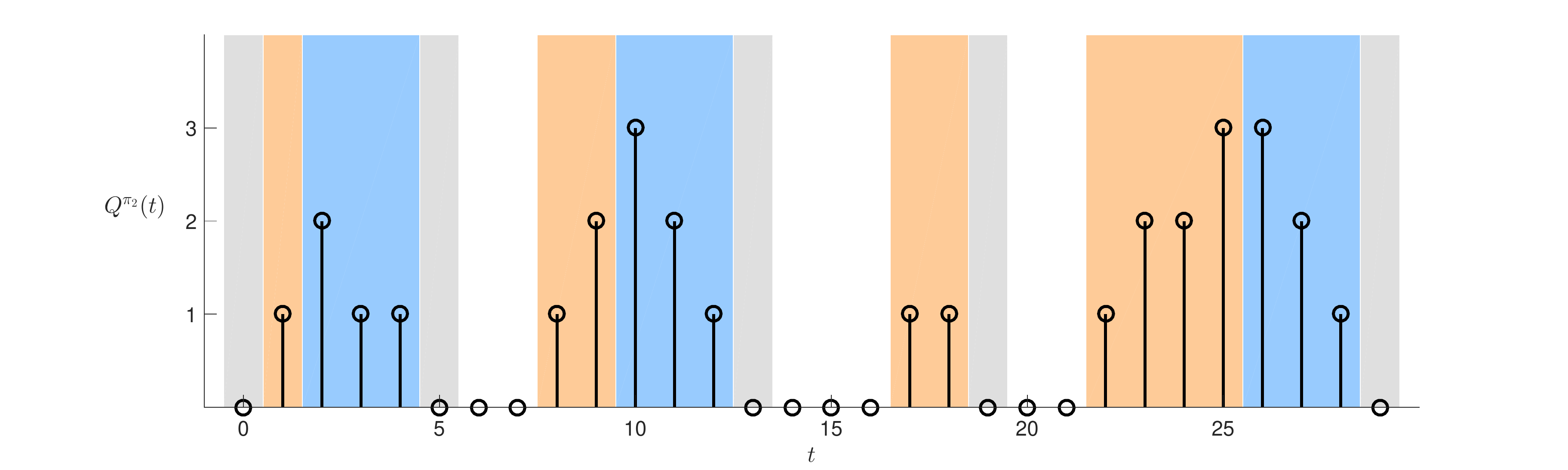}
\end{center}
  \caption{Example sample path of $Q^{\pi_2}(t)$ for some, unspecified, $(\lambda, \vec{\mu})$. The background colors indicate the actions of policy $\pi_2$  (see Fig.~\ref{policy_assumption_2}).  During time slots marked in gray, $\pi_2$ randomly updated one of the sample means, $\widehat{\mu}_i$.  During time slots marked in orange, $\pi_2$ scheduled the server with the (so-far) largest sample mean.  During time slots marked in blue, $\pi_2$ used the randomized policy defined by $\alpha_i$, because it had crossed the corresponding busy period's time-out threshold.  Note that the threshold increases linearly with each busy period number, and, for this example, the threshold was not reached in the third busy period.}
  \label{pi_2_illustrated_figure}
\end{figure}

A policy that achieves Theorem~\ref{theorem2} is given in Fig.~\ref{policy_assumption_2} and is referred to as $\pi_2$ throughout this subsection.  The policy, similar to the policy $\pi_1$ of Subsection~\ref{section_1_1}, iterates over empty and busy periods.  See Fig.~\ref{pi_2_illustrated_figure} for an illustration of the policy.  As with policy $\pi_1$, in the first time slot of each empty period, the policy schedules one server uniformly at random, observes the resulting offered service, and uses the observation to update a sample mean that estimates the server's service rate.  Again as with policy $\pi_1$, $\pi_2$ only uses the observations made during these time slots to update its sample means and does not use observations made during other time slots.

At the start of each busy period $p$, policy $\pi_2$ starts the busy period by scheduling the server with the highest sample mean for the first $p$ time slots of the busy period or until the queue empties and ends the busy period (whichever occurs first).  However, if the busy period lasts longer than $p$ time slots the policy hits a time-out threshold and resorts to using the randomized policy defined by Assumption~\ref{assumption2} to bring the queue backlog back to the empty state.  The time-out threshold grows linearly in busy period number $p$, and therefore with each busy period, the controller becomes more reluctant to call upon the randomized policy.  In the following analysis, we will see that this growing reluctance is important to establishing the proof of Theorem~\ref{theorem2}.

We now establish Theorem~\ref{theorem2} by analyzing $\pi_2$ on an arbitrary $(\lambda, \vec{\mu}) \in \mathcal{P}_2(\vec{\alpha})$. The proof is derived through three lemmas.  We begin by introducing some new notation that will facilitate understanding.

We begin by introducing notation that allows us to reference the period number $p$ that a time slot $t$ corresponds to.
 To this end, for each time $t$, define $p(t)$ to be the period number for the empty or busy period in which $t$ resides.  Note that for a given sample path, $p(t)$ maps time slots $t = 0, 1, \dots$ to their corresponding period number $p = 1, 2, \dots$.  For example, in Fig.~\ref{pi_2_illustrated_figure}, for $t = 0, 1, \dots, 4$, $p(t) = 1$ since these time slots occur during the first empty and busy periods and for $t = 5, 6, \dots, 12$, $p(t) = 2$ since these time slots occur during the second empty and busy periods.

Now, in our following analysis, we will be concerned with identifying the busy periods in which the policy does not strictly schedule server $i^*$ over the entire busy period.  Note that an ideal policy would schedule $i^*$ over all busy periods, and therefore we wish to identify during which busy periods policy $\pi_2$ falls short of this ideal.  To facilitate this analysis, for each busy period $p$, we let $C(p) \in \{0, 1\}$ be an indicator that takes the value $1$ if: either a server not equal to $i^*$ is first scheduled by policy $\pi_2$ at the start of the busy period or the time-out threshold is hit.  Note that by hitting the time-out threshold, we mean that the queue did not empty during the first $p$ time slots of busy period $p$.  Otherwise, if neither of these two conditions are met, we let the indicator $C(p) = 0$.  Note that by the definition of $\pi_2$, if any server $i \neq i^*$ is scheduled during busy period $p$, then $C(p)$ must equal $1$ (see Fig.~\ref{policy_assumption_2}). This is because, by the definition of policy $\pi_2$, we can only schedule a server $i \neq i^*$ during a busy period if: either the policy began the busy period by scheduling $i \neq i^*$ or it began the busy period scheduling server $i^*$ but still hit the time-out threshold, leading it to schedule the stabilizing policy.  Since to minimize regret, our objective is to mimic as best as we can the policy that schedules $i^*$ for all time slots, we can notionally view $C(p)$ equaling $1$ as policy $\pi_2$ failing to meet this ideal over busy period $p$.

In the following, we will combine the above two notations as $C(p(t))$.  Then, if $t$ is in a busy period, $C(p(t)) = 1$ if a server not equal to $i^*$ is scheduled at any time during the busy period in which $t$ belongs.  If $t$ is in an empty period, $C(p(t))$ indicates whether during the busy period following this empty period, a server not equal to $i^*$ is ever scheduled.  Then,
\begin{equation*}
\sum_{t = 0}^{T - 1} Q^{\pi_2}(t) C\left(p(t)\right)
\end{equation*}
is the queue backlog integrated, up to time slot $T$, over those busy periods for which the conditions of $C(p)$ are met.  Note that if $t$ is in an empty period, $Q^{\pi_2}(t) = 0$, and therefore the time slot contributes a value of $0$ to the above summation regardless of the value of $C\left(p(t)\right)$.  For example, for the sample path in Fig.~\ref{pi_2_illustrated_figure}, if $\pi_2$ scheduled $i^*$ for the start of busy period $3$, then $C(1) = C(2) = C(4) = 1$ since these busy periods reached their time-out thresholds, $C(3) = 0$, and $\sum_{t = 0}^{29} Q^{\pi_2}(t) C\left(p(t)\right) = 28$.  Likewise, if for the sample path, $\pi_2$ did not schedule $i^*$ for the start of busy period $3$, then $C(1) = C(2) = C(3) = C(4) = 1$ and $\sum_{t = 0}^{29} Q^{\pi_2}(t) C\left(p(t)\right) = 30$.

Given this notation, we have the following lemma, which states that the regret is upper bounded in expectation by the sum of queue backlogs over those busy periods in which indicator $C(p) = 1$.  The proof is similar to that of Lemma~\ref{lemma1} and can be found in Appendix~\ref{lemma5_appendix}.
\begin{lemma}
\label{lemma5}
$R_{(\lambda, \vec{\mu})}^{\pi_2}(T) \leq E \left[ \sum_{t = 0}^{T - 1} Q^{\pi_2}(t) C\left(p(t)\right) \right]$.
\end{lemma}

Now, to upper bound the right-hand side of Lemma~\ref{lemma5}, we will want to bound the probability that indicator $C(p) = 1$.  Recall that $C(p)$ can equal $1$ if one of the following events occurs: the policy begins the busy period by choosing a server other than $i^*$ to be scheduled for the first $p$ time slots or if it crosses the time-out threshold.  In the following lemma we establish that the probability of either event occurring diminishes exponentially in $p$.  To understand the intuition behind this result, first note that at each busy period $p$, the probability that there exists a server $i \neq i^*$ with a $\widehat{\mu}_i \geq \widehat{\mu}_{i^*}$ decreases exponential with $p$. This follows from the fact that the policy updates one of the sample means with a new observation at each empty period and Hoeffding's inequality.  Thus, the probability that at busy period $p$, $i^*$ is not chosen to be scheduled over the first $p$ time slots, is also exponentially decreasing with $p$.  Note that this argument is very similar to the one made in the proof of Lemma~\ref{lemma4}.  Now suppose that $i^*$ is scheduled at the start of busy period $p$. Then, since $\mu_{i^*} > \lambda$, we can show that the probability the queue does not empty before reaching the time-out threshold $p$ must decrease exponentially in $p$ as well.  We now state the lemma whose proof can be found in Appendix~\ref{lemma6_appendix}.
\begin{lemma}
\label{lemma6}
There exists positive constants $M_0$, $\chi$, and $p_0$ such that for all $p \geq p_0$,
\begin{equation*}
P\left( C(p) = 1 \right) \leq M_0 e^{-\chi p}.
\end{equation*}
\end{lemma}

To complete our upper bound on the right-hand side of Lemma~\ref{lemma5}, we will need to bound the cost of indicator $C(p)$ equaling $1$.
To this end, let variable $Z^{\pi_2}(p) \in \left\{1, 2, \dots \right\}$ denote the integral of the queue backlog taken over busy period $p$.  For example, for the sample path shown in Fig.~\ref{pi_2_illustrated_figure}, $Z^{\pi_2}(1) = 5$ and $Z^{\pi_2}(2) = 9$.  In Lemma~\ref{lemma7}, we show $E\left[ \left. Z^{\pi_2}(p) \right \rvert C(p) = 1 \right]$ grows at most quadratically with $p$.  This is because the queue backlog can grow at most linearly over the first $p$ time slots of the busy period.  Thus, if the time-out threshold is hit, the required amount of time that the randomized policy will need to empty the queue can also be at most on the order of $p$.  Since, the maximum queue backlog at any time slot in a given busy period cannot be greater than the busy period's duration, the expected integrated queue backlog will at most be on the order of $p^2$.
\begin{lemma}
\label{lemma7}
There exists positive constants $M_1$ and $\beta_1$ such that for all $p$,
\begin{equation*}
E\left[ \left. Z^{\pi_2}(p) \right \rvert C(p) = 1 \right] \leq M_1 p^2 + \beta_1.
\end{equation*}
\end{lemma}
The proof of the lemma can be found in Appendix~\ref{lemma7_appendix}.

Before continuing to the proof of the theorem, we observe that $E\left[ \left. Z^{\pi_2}(p) \right \rvert C(p) = 0 \right] \leq p^2$, since $C(p) = 0$ implies that the time-out threshold was not reached.  Thus, we see that for all $p$, $E\left[Z^{\pi_2}(p) \right]$ is finite (implying the expected duration of each busy period is finite under policy $\pi_2$).  We are now ready to establish the theorem.

\begin{IEEEproof}[Proof of Theorem~\ref{theorem2}]
Beginning with Lemma~\ref{lemma5}, we have that the regret is bounded by the integrated queue backlog over those busy periods such that the conditions of $C(p)$ are met. i.e., 
\begin{equation}
\label{bounding_regret_by_events1}
R_{(\lambda, \vec{\mu})}^{\pi_2}(T) \leq E \left[ \sum_{t = 0}^{T - 1} Q^{\pi_2}(t) C\left(p(t)\right) \right].
\end{equation}

Now, no more than $T$ busy periods can occur in $T$ time slots, since each busy period, by definition, must be one time slot long.  In fact, as was noted in the previous subsection, there must be much fewer than $T$ busy periods in $T$ time slots, since the empty periods in between the busy periods must also be at least one time slot long, as well.  Thus, we can bound the right hand side of  \eqref{bounding_regret_by_events1}, which is the integral of the queue backlogs over busy periods with $C(p) = 1$ up to time $T$, with the integral of the queue backlogs over busy periods with $C(p) = 1$ through the $T^{th}$ busy period.  Formally,
\begin{equation}
\eqref{bounding_regret_by_events1} \leq E \left[ \sum_{p = 1}^T Z^{\pi_2}(p) C(p) \right].
\label{bounding_regret_by_events2}
\end{equation}
Note that in going from \eqref{bounding_regret_by_events1} to \eqref{bounding_regret_by_events2}, we have changed from summing over time slots $t$ to summing over busy period number $p$.

Now, recall that $C(p)$ is an indicator.  When $C(p) = 0$, $Z^{\pi_2}(p) C(p) = 0$. Thus,
\begin{align}
\eqref{bounding_regret_by_events2}& = \sum_{p = 1}^T P(C(p) = 1) E\left[\left. Z^{\pi_2}(p) \right \rvert C(p) = 1 \right] \nonumber \\
& \leq \sum_{p=1}^{p_0 - 1} \left(M_1 p^2 + \beta_1\right) + \sum_{p=p_0}^T \left(M_1 p^2 + \beta_1\right) M_0 e^{-\chi p} \label{bounding_regret_by_events3}
\end{align}
where we have applied Lemmas~\ref{lemma6}~and~\ref{lemma7} to obtain the inequality.
Recall that $p_0$ is a constant and therefore the first term in \eqref{bounding_regret_by_events3} is bounded.  Likewise, for the second term in \eqref{bounding_regret_by_events3}
\begin{equation*}
\sum_{p=0}^\infty \left(M_1 p^2 + \beta_1\right) M_0 e^{-\chi p} < \infty,
\end{equation*}
which follows from well-known results for series.
 Thus, we see that $R_{(\lambda, \vec{\mu})}^{\pi_2}(T) = O(1)$, which gives the result.
\end{IEEEproof}

\subsection{Learning Stabilizing Policies}
\label{section_1_3}

We now relax Assumption~\ref{assumption2} and no longer assume that a randomized policy is given to the controller that can be relied upon to stabilize the queue.  As a result, the controller will have to learn how to take actions so as to empty the queue infinitely often.  Once the controller can accomplish this, it may use the empty periods to identify the best server.  To this end, we only make the following assumption about the parameters $\lambda$ and $\vec{\mu}$. 
\begin{assumption}
\label{assumption3} 
$(\lambda, \vec{\mu}) \in \mathcal{P}_3 \triangleq \left\{ \mathcal{P} : \mu_{i^*} > \lambda \right\}.$
\end{assumption}

We then have the following theorem, which is analogous to the previous subsections.  The rest of this subsection will be devoted to proving its statement.

\begin{figure}
\noindent\makebox[\linewidth]{\rule{\columnwidth}{0.4pt}}
\begin{algorithmic}
\STATE{ $\widehat{\mu}_i = 0$ for all $i \in [N]$}
\WHILE{TRUE}
	\FOR{Empty period $p$}
		\STATE{At the first time slot of the period, set $u(t) = i$ uniformly at random over $[N]$ and update $\widehat{\mu}_i$ with the observed state of $D(t)$}
		\STATE{For other time slots in the empty period, idle}
	\ENDFOR
	\FOR{Busy period $p$}
		\STATE{Schedule $\arg \max_{i \in [N]} \widehat{\mu}_i$ for the first $p$ time slots of the busy period or until the queue empties }
		\IF{The queue does not empty during the first $p$ time slots}
			\STATE{Use the algorithm of Fig.~\ref{policy_learning} to empty the queue}
		\ENDIF
	\ENDFOR
\ENDWHILE
\end{algorithmic}
\noindent\makebox[\linewidth]{\rule{\columnwidth}{0.4pt}}
\caption{The policy $\pi_3$ for achieving Theorem~\ref{theorem3}. Each variable $\widehat{\mu}_i$ is initialized to zero prior to its first update.}  The method uses the algorithm of Fig.~\ref{policy_learning} when the time-out threshold is hit.
\label{policy_assumption_3}
\end{figure}

\begin{figure}
\noindent\makebox[\linewidth]{\rule{\columnwidth}{0.4pt}}
\begin{algorithmic}
\STATE{ $\widehat{\mu}^p_i = 0$ for all $i \in [N]$}
\STATE{$V(0) = 0$}
\FOR{Time slot $n = 0, 1, \dots$ until queue empties}
	\IF{$n$ is a dedicated exploration time slot}
		\STATE{Choose $u(n) = i$ uniformly at random over $[N]$}
		\STATE{Update $\widehat{\mu}^p_i$ with the observed state of $D(n)$}
		\STATE{$V(n+1) = V(n) + 1$}
	\ELSE
		\STATE{Schedule $u(n) = \arg \max_{i \in [N]} \widehat{\mu}^p_i$}
		\STATE{$V(n+1) = V(n)$}
	\ENDIF
\ENDFOR
\end{algorithmic}
\noindent\makebox[\linewidth]{\rule{\columnwidth}{0.4pt}}
\caption{Learning algorithm for bringing the queue back to the empty state. Variable $n$ counts the number of time slots since the algorithm's invocation.  The method uses a new set of sample means $\widehat{\mu}^p_i$ to estimate the server rates.  Note that $\widehat{\mu}^p_i$ is a different variable from $\widehat{\mu}_i$, and $\widehat{\mu}^p_i$ is only updated using observations made during busy period $p$.  Each variable $\widehat{\mu}^p_i$ is initialized to zero.  The array $V(n)$ counts the number of dedicated exploration time slots that have occurred since the start of the algorithm's invocation.  We require that the locations of the dedicated exploration time slots be chosen such that $V(n)$ meets the condition of \eqref{sublinear_exploration} for each time slot $n$.}
\label{policy_learning}
\end{figure}

\begin{theorem}
\label{theorem3}
Under Assumption~\ref{assumption3}, $\exists \pi \in \Pi$ such that, for each $(\lambda, \vec{\mu}) \in \mathcal{P}_3$, $R_{(\lambda, \vec{\mu})}^\pi(T) = O(1)$.
\end{theorem}

To prove the theorem, we will analyze policy $\pi_3$ on an arbitrary $(\lambda, \vec{\mu}) \in \mathcal{P}_3$.  Policy $\pi_3$ is shown in Fig.~\ref{policy_assumption_3}.  The policy is very similar to $\pi_2$, except now when the time-out threshold of a busy period is reached, rather than relying on the known randomized policy given by Assumption~\ref{assumption2}, $\pi_3$ uses the method of Fig.~\ref{policy_learning} to bring the queue back to the empty state.  Starting from the time-out threshold's crossing, the method is run until the queue finally empties and the busy period ends.  As can be seen in Fig.~\ref{policy_learning}, we use time slot counter $n = 0, 1, \dots$ to indicate the number of time slots since the method's invocation (i.e., $n$ is the time slot number normalized to the time-out threshold's crossing).

In its execution, the method in Fig.~\ref{policy_learning} dedicates certain time slots to exploration. During these dedicated exploration time slots, the policy chooses from the servers uniformly at random, observes the offered service of the chosen server, and updates the corresponding sample mean $\widehat{\mu}^p_i$ using the observation.  Note that the sample means $\widehat{\mu}^p_i$ are new variables that are used by the method during its execution at busy period $p$ and are completely separate from the sample means $\widehat{\mu}_i$ in Fig.~\ref{policy_assumption_3}.  The sample means $\widehat{\mu}^p_i$ are initialized to zero at the start of the method, when the time-out threshold is crossed, and do not make use of any of the previous observations that form sample means $\widehat{\mu}_i$.  Likewise, sample means $\widehat{\mu}_i$ are not updated with any of the observations made during the method's dedicated exploration time slots.  Finally, it is worth stressing that each call to the method of Fig.~\ref{policy_learning}, at a new busy period $p$, initializes new sample means $\widehat{\mu}^p_i$ that are used only during this execution and do not share observations with previous executions of the method.  Thus, the variables $\widehat{\mu}^p_i$ are estimates of the sample means formed only by observations made during the exploration times slots in busy period $p$.  This property will be important to our later analysis.

Now, examining Fig.~\ref{policy_learning}, one may note that the exact locations of the dedicated exploration time slots are not specified.  In fact, we will allow for an implementer of the method to choose their exact locations subject to some constraints that we now specify.  We assume the locations are fixed and are specified relative to the start of the method's invocation.  We also require that the same relative locations be used for all busy periods.  Thus, the implementer must choose for which values of $n$ in Fig.~\ref{policy_learning} dedicated explorations taken place and must use the same choice at each busy period that the method is invoked.  To give an example, one choice could be to have dedicated explorations occur at times $n = j^2$ for $j = 0, 1, 2, 3, 4,  \dots$ (i.e., $n = 0, 1, 4, 9, 16, \dots$).

We now specify one last requirement for the design of the exploration time slots' locations.  Let $V(n)$ be the number of dedicated explorations that are designed to occur during the method's first $n$ time slots.  (For illustration, in Fig.~\ref{policy_learning}, we show the method explicitly counting $V(n)$.)  Then for our proof of Theorem~\ref{theorem3}, we require that the dedicated exploration times are chosen such that
\begin{equation}
\label{sublinear_exploration}
r n^\epsilon - b_1 \leq V(n) \leq r n^\epsilon + b_2, \mbox{ for } n = 0, 1, 2, \dots
\end{equation}
for some positive constants $b_1$, $b_2$, $r$, and $\epsilon \in (0, 1)$.
Note that \eqref{sublinear_exploration} requires that the frequency of explorations diminishes with $n$ and eventually falls below $1 - \frac{\lambda}{\mu_{i^*}}$.  Subject to \eqref{sublinear_exploration}, the exact choice of the dedicated exploration times is left to implementation.  Observe that our previous example that chose locations $n = j^2$ meets the condition of \eqref{sublinear_exploration}.

We now turn to establishing the proof of Theorem~\ref{theorem3}.  Since policy $\pi_3$ is very similar to policy $\pi_2$, the proof of Theorem~\ref{theorem3} will closely follow the proof of Theorem~\ref{theorem2}.  The main difference will be that we use Lemma~\ref{lemma8} below in place of Lemma~\ref{lemma7} of the previous subsection.

We begin by noting that, using the definitions of $p(t)$ and $C(p)$ from the previous subsection, Lemmas~\ref{lemma5}~and~\ref{lemma6} continue to hold for policy $\pi_3$.  Specifically, as before, if we schedule a server $i \neq i^*$ during any time slot of a busy period $p$, than we must either have started the busy period by scheduling a server $i \neq i^*$ or we must have hit the time-out threshold.  By the arguments of the previous subsection for policy $\pi_2$, both events again diminish exponentially in $p$ for policy $\pi_3$.  Then, the only step that is really missing in establishing the theorem is bounding the cost of having $C(p) = 1$ for busy period $p$.  We proceed to do this next.

For policy $\pi_3$, let $Z^{\pi_3}(p) \in \{1, 2, \dots\}$ denote the integrated queue backlog of busy period $p$.  Then, we have the following lemma whose proof can be found in Appendix~\ref{lemma8_appendix}. 
\begin{lemma}
\label{lemma8}
There exists positive constants $M_2$ and $\beta_2$ such that for all $p$,
\begin{equation*}
E\left[ \left. Z^{\pi_3}(p) \right \rvert C(p) = 1 \right] \leq M_2 p^2 + \beta_2.
\end{equation*}
\end{lemma}

Given Lemma~\ref{lemma8}, we now establish the proof.  Note that, due to its similarity to the proof of Theorem~\ref{theorem2}, we have abridged a few steps.
\begin{IEEEproof}[Proof of Theorem~\ref{theorem3}]
Using Lemma~\ref{lemma5} and the fact that no more than $T$ busy periods can occur in $T$ time slots, 
\begin{equation}
\label{bounding_regret_by_events_theorem3}
R_{(\lambda, \vec{\mu})}^{\pi_3}(T) \leq E \left[ \sum_{t = 0}^{T - 1} Q^{\pi_3}(t) C\left(p(t)\right) \right] \leq E \left[ \sum_{p = 1}^T Z^{\pi_3}(p) C(p) \right].
\end{equation}
Then, as in the proof of Theorem~\ref{theorem2}, we may write
\begin{align*}
\eqref{bounding_regret_by_events_theorem3}& = \sum_{p = 1}^T P(C(p) = 1) E\left[\left. Z^{\pi_3}(p) \right \rvert C(p) = 1 \right] \nonumber \\
& \leq \sum_{p=1}^{p_0 - 1} \left(M_2 p^2 + \beta_2\right) + \sum_{p=p_0}^T \left(M_2 p^2 + \beta_2\right) M_0 e^{-\chi p}
\end{align*}
where the last inequality follows from Lemmas~\ref{lemma6}~and~\ref{lemma8}.
Since 
\begin{equation*}
\sum_{p=0}^\infty \left(M_2 p^2 + \beta_2\right) M_0 e^{-\chi p} < \infty,
\end{equation*}
 $R_{(\lambda, \vec{\mu})}^{\pi_3}(T) = O(1)$.
\end{IEEEproof}

\section{Systems without Stabilizing Servers}
\label{section_1_4}

In the previous section, we saw that the existence of a stabilizing server (i.e., $\mu_{i^*} > \lambda$) allowed for policies that had $R_{(\lambda, \vec{\mu})}^\pi(T) = O(1)$.  A natural question is whether we can obtain a similar result for the subset of problems $(\lambda, \vec{\mu})$ that do not have stabilizing servers.  We proceed to show that there cannot exist a policy that can achieve $O(1)$ regret over this entire subset.  The intuition behind this result is that, when $\mu_{i^*} \leq \lambda$, the system may only experience a finite number of empty periods, and therefore, policies cannot rely upon the empty periods to perform sufficient exploration of the servers' rates.

Let $\mathcal{P}_4 \triangleq \left\{\mathcal{P} : \mu_i \leq \lambda, \forall i \in [N] \right\}$. 
Then, we have the following theorem.
\begin{theorem}
\label{theorem4}
For any policy $\pi \in \Pi$ there exists a $(\lambda, \vec{\mu}) \in \mathcal{P}_4$ such that $R_{(\lambda, \vec{\mu})}^\pi(T) = \Omega(T)$.
\end{theorem}

\begin{IEEEproof}
Consider a system with $\lambda = 1$.  Let $\pi$ be any arbitrary learning policy in $\Pi$.  Then for all $t > 0$ the queue is non-empty.  As a result, the queue backlogs are given by
\begin{align}
Q^\pi(t) &= t - \sum_{\tau = 1}^{t-1} D^\pi(\tau) \label{queue_backlog_under_lambda_equals_1_1} \\
Q^*(t) &= t - \sum_{\tau = 1}^{t-1} D^*(\tau) \label{queue_backlog_under_lambda_equals_1_2} 
\end{align}
where $D^*(t)$ is the service offered by server $i^*$ and $D^\pi(t)$ is the service offered by the server scheduled by $\pi$ at time $t$.
By the above equations, it is clear that at $t = 0$, $Q^\pi(0) = Q^*(0) = 0$ and at $t = 1$, $Q^\pi(1) = Q^*(1) = 1$.  Thus, $R_{(\lambda, \vec{\mu})}^\pi(0) = R_{(\lambda, \vec{\mu})}^\pi(1) = 0$.  For $T > 1$, equations \eqref{queue_backlog_under_lambda_equals_1_1} and \eqref{queue_backlog_under_lambda_equals_1_2} imply
\begin{equation}
\label{linearly_growing_regret}
R_{(\lambda, \vec{\mu})}^\pi(T) = \sum_{t=2}^T E\left[ \sum_{\tau=1}^{t-1} D^*(\tau) - \sum_{\tau=1}^{t-1} D^\pi(\tau) \right].
\end{equation}
And, we can write
\begin{equation}
\label{regret_bounded_by_bad_pulls}
E\left[ \sum_{\tau=1}^{t-1} D^*(\tau) - \sum_{\tau=1}^{t-1} D^\pi(\tau) \right] = \sum_{i \in [N] - i^*} \left(\mu_{i^*} - \mu_i\right) E\left[\sum_{\tau = 1}^{t-1} \mathbf{1}\left\{ u(\tau) = i \right\}\right]
\end{equation}
where $u(\tau)$ denotes the decision made by policy $\pi$ at time $\tau$.

Now, for some $\vec{\mu}$ such that $(1, \vec{\mu}) \in \mathcal{P}_4$, there exists a $\delta > 0$ such that for some $i \neq i^*$, 
\begin{equation}
\label{perf_bounded_from_zero}
E\left[\mathbf{1}\left\{ u(1) = i \right\}\right] \geq \delta.
\end{equation}
Otherwise, there would exist a learning policy that, given any $\vec{\mu}$, always schedules the optimal server with probability one at time slot $1$.  (This would be a policy that guesses the best server with probability one, having not even made an observation of every server prior to its guess.)  Clearly, no such policy in $\Pi$ can exist.  Therefore, because $E\left[\sum_{\tau = 1}^{t-1} \mathbf{1}\left\{ u(\tau) = i \right\}\right]$ is monotonically increasing in $t$, \eqref{perf_bounded_from_zero} implies that
\begin{equation*}
E\left[\sum_{\tau = 1}^{t-1} \mathbf{1}\left\{ u(\tau) = i \right\}\right] \geq \delta
\end{equation*}
for all $t \geq 2$ and some $\vec{\mu}$.
Using this in \eqref{regret_bounded_by_bad_pulls} and then \eqref{linearly_growing_regret} gives the result.
\end{IEEEproof}

\section{Simulation Results}
\label{section_simulation}

\begin{figure}

\noindent\makebox[\linewidth]{\rule{\columnwidth}{0.4pt}}
\begin{algorithmic}
\STATE{For the first $N$ time slots, sample the servers in order}
\WHILE{TRUE}
	\FOR{Empty period $p$}
		\STATE{At each time $t$ in the empty period, use the empty-period sampling rule to choose $u(t)$ and update $\widehat{\mu}_{u(t)}$ with the observed state of $D(t)$}
	\ENDFOR
	\FOR{Busy period $p$}
		\STATE{For the first $p$ time slots of the busy period or until the queue empties, schedule $u(t) = \arg \max_{i \in [N]} \widehat{\mu}_i$, updating $\widehat{\mu}_{u(t)}$ with the observation of $D(t)$ }
		\IF{The queue does not empty during the first $p$ time slots}
			\STATE{At each time $t$ until the queue empties, schedule the server with the maximum UCB1 weight, using the observed state $D(t)$ to update $\widehat{\mu}_{u(t)}$ }
		\ENDIF
	\ENDFOR
\ENDWHILE
\end{algorithmic}
\noindent\makebox[\linewidth]{\rule{\columnwidth}{0.4pt}}
\caption{The heuristic technique used in our simulations. The method maintains sample mean variables $\widehat{\mu}_i$ for each server, which are updated with the observed offered service every time that server is chosen.  We consider three different rules for sampling servers during empty periods.  Arrivals to the queue begin after time step $N$.}
\label{simulation_policy}
\end{figure}

The analysis of Section~\ref{section_1} considered policies that proved the achievablility of $O(1)$ queue length regret.  As discussed in Section~\ref{section_0} these methods cause the tail of the regret to trend to zero.  In order to keep our analysis tractable, the designed policies had to make concessions, described further below, that hurt their practical performance.  Nevertheless, these policies do provide intuition about the problem, and with some fairly minor and intuitive changes, we show that these policies can be transformed into heuristic techniques that achieve good performance in simulation.  The heuristics considered in this section use policy $\pi_3$ as a template.

We now briefly cover the inefficiencies of the policy $\pi_3$ of Section~\ref{section_1_4}.  The first, and easiest to remedy, is that the policy only uses the first time slot of each empty period to sample a server.  Clearly, a more efficient method would sample throughout the empty period, and the only reason policy $\pi_3$ was designed not to do this is because it is not strictly necessary for the proof of Theorem~\ref{theorem3} and simplifies the mathematics.

The next inefficiency is that policy $\pi_3$ does not record the observations made of the chosen server during busy periods prior to reaching the threshold, and, after reaching the threshold, does not use its observations after the busy period expires.  The reason for this is to facilitate simplicity in the proof of Theorem~\ref{theorem3}.  Observations made during the busy period are correlated with the queue's dynamics, and therefore if these observations are used during subsequent busy periods, one has to account for how the dynamics impact the probabilities of the previous observations.  This would greatly complicate analysis.  However, in practice, it makes sense that a policy should use all previous observations to inform its estimates of the servers' rates.

Finally, policy $\pi_3$ uses a simple exploration process for identifying a server to empty the queue once the threshold is reached during a busy period.  This method is relatively easy to analyze with the queue's dynamics.  However, algorithms from the literature on multi-armed bandits, such as UCB1 described below, are generally better methods and should probably be used in practice.  Analyzing how the mechanisms used by these algorithms to achieve good expected performance interact with the queue's dynamics to reach the next empty period is an interesting direction for future research.    

Given the above points, in Fig.~\ref{simulation_policy} we provide a heuristic technique that uses the intuition provided in the analysis of policy $\pi_3$.  The technique uses one of three methods to explore the servers during empty periods.  Method UCB-LE chooses to sample at each time step in the empty period the server that has been least explored so-far (i.e., observed the least number of times).  Method UCB-UE randomly chooses a server to explore at each time step of the empty period, uniformly choosing among the servers.  Finally, method UCB-WE randomly chooses from the servers proportional to the value of $\widehat{\mu}_i + b$, where $\widehat{\mu}_i$ is the sample mean of the observations made so-far of server $i$ and $b$ is an additional term (set to $0.1$ in our simulations).  UCB-WE weights its explorations during empty periods towards the best performing servers.

\begin{figure*}
    \centering
  \subfloat[]{%
        \includegraphics[width=0.45\linewidth]{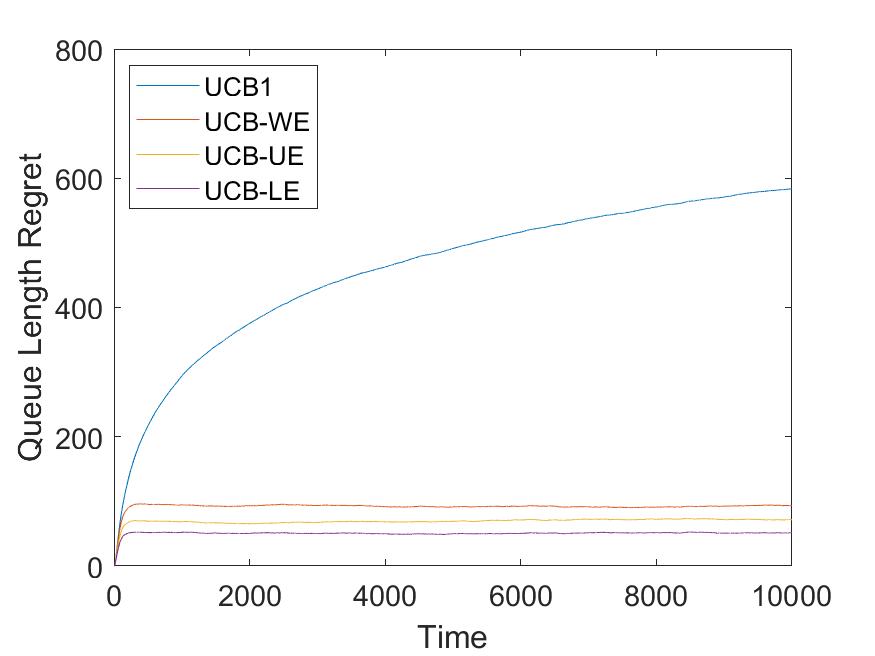}}
  \subfloat[]{%
        \includegraphics[width=0.45\linewidth]{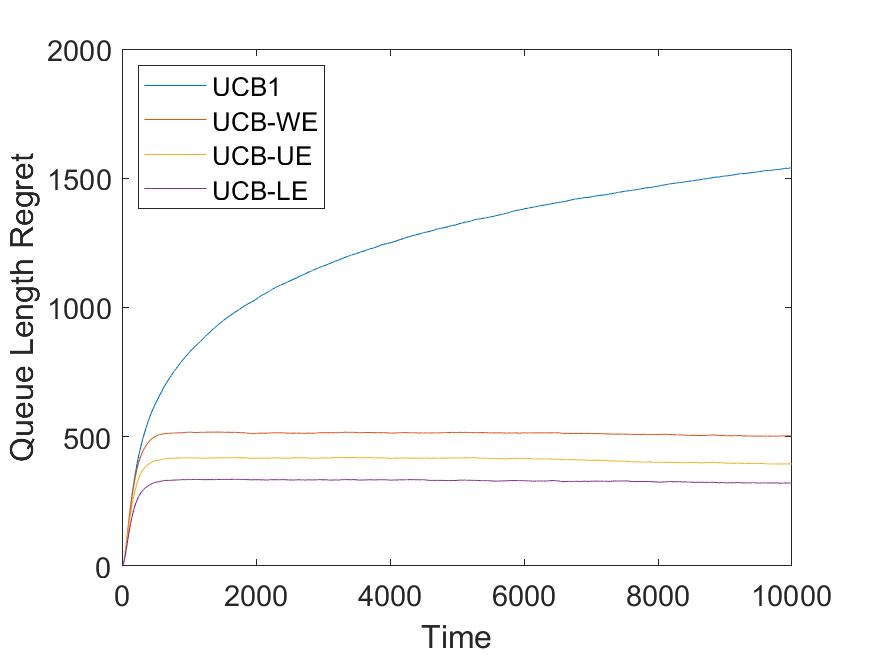}}
        \\
  \subfloat[]{%
        \includegraphics[width=0.45\linewidth]{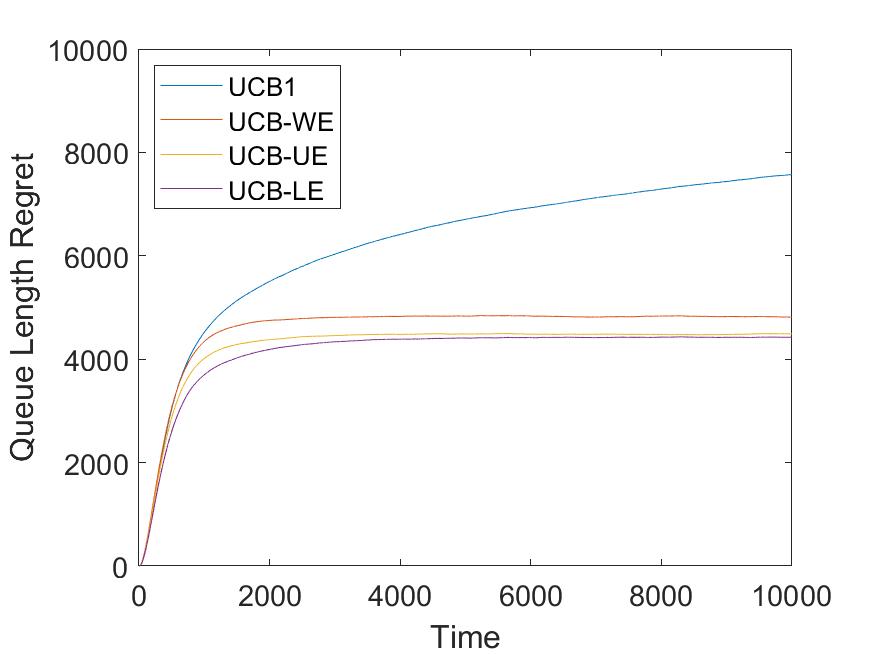}}
  \caption{Simulations of a four server problem with service rates $\vec{\mu} = (0.1, 0.3, 0.5, 0.7)$. Different arrival rates $\lambda$ are considered. (a) $\lambda = 0.4$  (b) $\lambda = 0.5$  (c) $\lambda = 0.6$. }
  \label{four_server_scenarios} 
\end{figure*}

During busy periods, the heuristic of Fig.~\ref{simulation_policy} begins by choosing to schedule at each time slot the server with the highest sample mean until the busy-period threshold is reached (or until the queue empties).  The observations made at this time are used to update the sample means of the servers.  Once the threshold is reached, the method constructs at each time a UCB1 weight for each server and schedules the server with the highest weight.  The UCB1 weight, proposed in \cite{auer} for solving the multi-armed bandit problem, is given by
\begin{equation*}
\widehat{\mu}_i + \sqrt{\frac{2 \ln(t + 1)}{T_i}}
\end{equation*}
where $\widehat{\mu}_i$ is the sample mean of the service that has been offered by server $i$ so-far and $T_i$ is the number of observations made of the server.  The method schedules according to the UCB1 weights until the queue empties.

In the following, we will compare the three versions of our heuristic (UCB-LE, UCB-UE, and UCB-WE) against a method that always chooses servers according to the UCB1 weight (during both empty and busy periods).  UCB1 is a popular multi-armed bandit algorithm \cite{auer}. In our context, it is a method that maximizes the amount of offered service to the queue without regard to the queue's state.  In our simulations, all methods will spend the first $N$ time slots sampling each of the $N$ servers once before launching into their main techniques.  The queue is kept empty during these first $N$ time slots, and arrivals to the queue begin afterwards.  In our plots, we calculate the queue length regret by averaging, across multiple runs, the difference in the queue backlog of the learning policies and the (genie) policy that always schedules the maximum rate server. 

In Fig.~\ref{four_server_scenarios}, we explore the impact of increasing arrival rates on the four methods' performance.  In this scenario there are four servers with service rates 0.1, 0.3, 0.5, and 0.7.  Arrival rates of 0.4, 0.5, and 0.6 are simulated.  The queue length regret averaged over 10,000 simulation runs is plotted versus time.  In the three plots, we see that the heuristic methods generally outperform UCB1 in minimizing queue backlog.  The queue length regret grows with increasing arrival rate as the system becomes more heavily loaded and there are fewer opportunities for free exploration during empty periods.  Additionally, as the arrival rate becomes greater, the amount of time available for the queue to be empty decreases, and the relative difference in performance between the heuristic methods and UCB1 decreases.  This effect appears to become more severe as the system becomes more heavily loaded.

\begin{figure*}
    \centering
  \subfloat[]{%
        \includegraphics[width=0.45\linewidth]{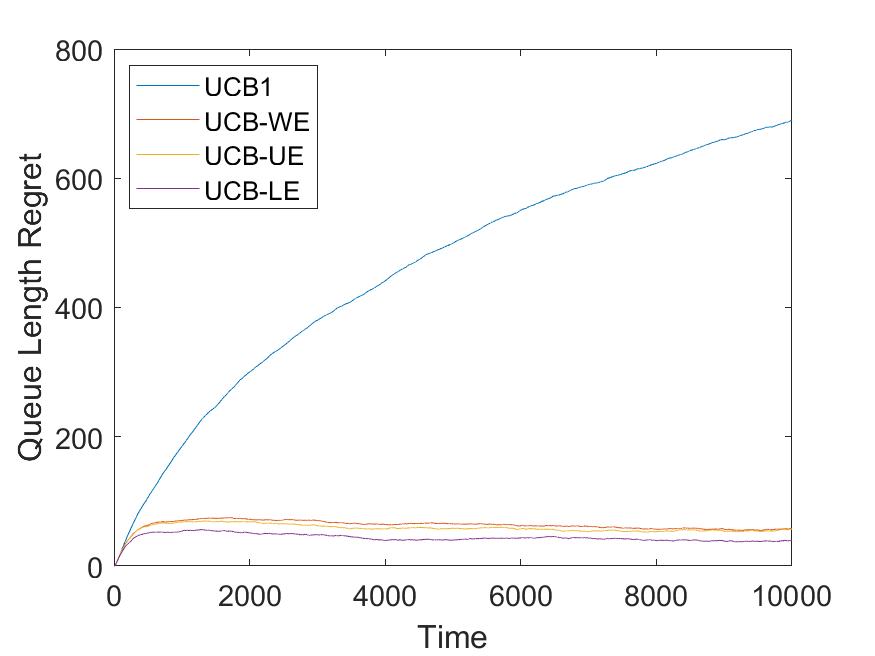}}
  \subfloat[]{%
        \includegraphics[width=0.45\linewidth]{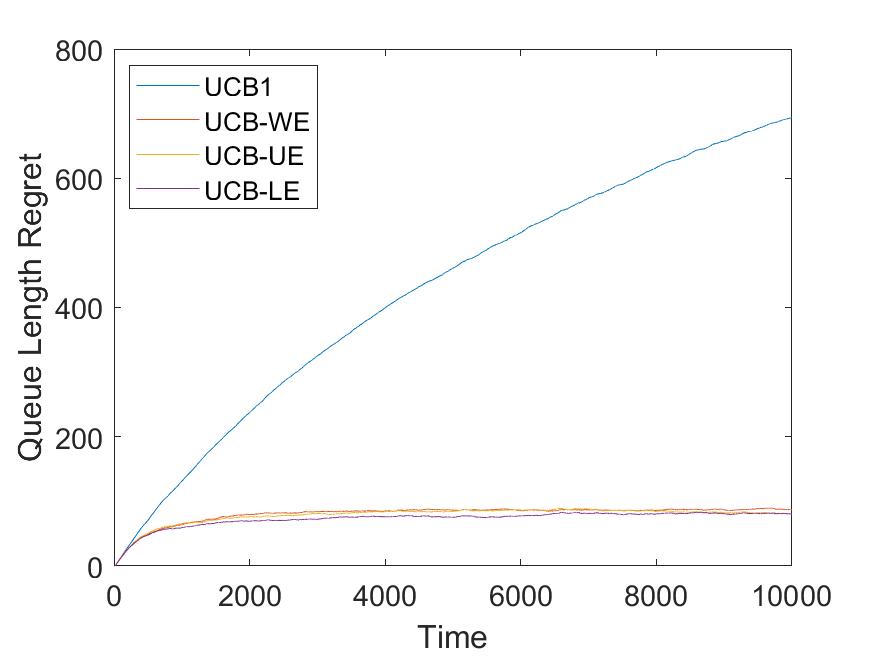}}
        \\
  \subfloat[]{%
        \includegraphics[width=0.45\linewidth]{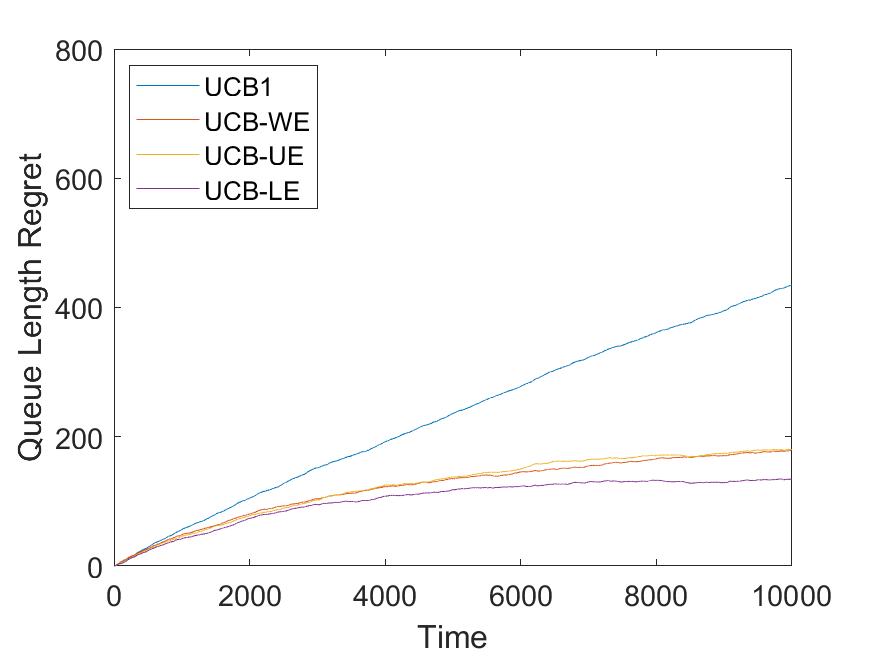}}
  \caption{Simulations of a two server problem with arrival rate $\lambda = 0.4$. Different service rates are considered. (a) $\vec{\mu} = (0.5, 0.6)$  (b) $\vec{\mu} = (0.54, 0.6)$  (c) $\vec{\mu} = (0.58, 0.6)$.}
  \label{two_server_scenarios} 
\end{figure*}

Fig.~\ref{two_server_scenarios} examines how a narrowing gap between server rates impacts performance.  In this scenario there are two servers.  The first server has its service rate tested at values of 0.5, 0.54, and 0.58, while the second server has a fixed rate of 0.6.  The arrival rate to the queue is 0.4 and each plot shows the queue length regret averaged over 10,000 simulation runs.  We see that decreasing the gap between the server rates causes the heuristic methods' queue length regret to converge more slowly as the methods have a more difficult time settling on the optimal server.  Although the methods outperform UCB1 in all three cases, their relative advantage, over the considered timescales, generally decreases as the gap between the servers' rates decreases.

\section{Conclusion}
\label{conclude}
In this work, we considered a queueing system that has $N$ available servers with unknown service rates.  The objective of the system controller is to minimize the queue's backlog by using observations of the service process to quickly identify the best server.  To evaluate a learning policy's performance, we introduced queue length regret, which is defined to be the expected difference in the accumulated queue backlog of the learning policy and the accumulated backlog of a controller that always schedules the optimal server.  This metric quantifies the cost, in queueing delay, of having to learn the best server.

We proved that there exist queue-length based policies that have an order optimal $O(1)$ regret for all systems such that $\mu_{i^*} > \lambda$.  These policies achieve this result by taking advantage of times when the queue is empty and  estimates of the servers can be improved without incurring extra packet delay.  Importantly, we showed there exist policies that can ensure the queue empties infinitely often, so that this property can be exploited.  In contrast to this result, we showed that traditional bandit algorithms that focus on only maximizing offered service to the system without accounting for the queue's state can have a queue length regret that is $\Omega(\log{T})$.  Thus, to obtain optimal performance, it is insufficient to design controllers that only focus on maximizing the offered service to the queue.  

In the problem considered herein, the controller's estimates of the servers' rates are highly correlated with the queue's dynamics, and therefore analyzing the system's stochastic behavior can be difficult.  The theoretical policies analyzed in this work were chosen for their analytical tractability, and the simplicity of the chosen policies allowed us to use well-known tools from statistics to show the achievability of $O(1)$ regret, which was our main objective.   We then used insight from our theoretical analysis to design heuristic techniques with good performance.  Finding policies with provable optimal performance over finite-time is a possible direction for future research.


%

\appendices


\section{Proof of Lemma~\ref{lemma1}}
\label{lemma1_appendix}

Consider an outcome $\omega$ and fixed sample paths $A(t, \omega)$ and $D^i(t, \omega)$, $\forall i \in [N]$.  Suppose at time $t$, policy $\pi_1$ is scheduling $i^*$ (i.e., $u(t, \omega) = i^*$) and $Q^{\pi_1}(t, \omega) > 0$.  Define, $s < t$ the last time before $t$ that the queue was empty (i.e., the time slot preceding the start of this busy period).  Then, since $\pi_1$ does not change servers mid-busy period,
\begin{equation*}
Q^{\pi_1}(t, \omega) = \sum_{\tau = s}^{t - 1} A(\tau, \omega) - \sum_{\tau = s + 1}^{t - 1} D^*(\tau, \omega),
\end{equation*}
where $D^*(\tau, \omega)$ is equal to the service offered by server $i^*$ at time $\tau$.  Now, let $Q^*(\tau, \omega)$ be the queue backlog under the controller that always schedules $i^*$. Since $\left(Q^*(s, \omega) - D^*(s, \omega)\right)^+ \geq 0$, this implies that
\begin{equation*}
Q^*(t, \omega) = \sum_{\tau = s}^{t - 1} A(\tau, \omega) - \sum_{\tau = s + 1}^{t - 1} D^*(\tau, \omega) + \left(Q^*(s, \omega) - D^*(s, \omega) \right)^+ \geq Q^{\pi_1}(t, \omega).
\end{equation*}

Summing over time we then obtain
\begin{equation*}
\sum_{t = 0}^{T-1} Q^{\pi_1}(t, \omega) \mathbf{1}\left\{ u(t, \omega) = i^* \right\} \leq \sum_{t = 0}^{T - 1} Q^*(t, \omega).
\end{equation*}
Taking expectation with respect to arrivals, departures, and the randomization in policy $\pi_1$,
\begin{equation*}
E \left[ \sum_{t = 0}^{T-1} Q^{\pi_1}(t) \mathbf{1}\left\{ u(t) = i^* \right\} \right] \leq E \left[ \sum_{t = 0}^{T - 1} Q^*(t) \right].
\end{equation*}
Then, starting with the definition of $R_{(\lambda, \vec{\mu})}^{\pi_1}(T)$,
\begin{align*}
R_{(\lambda, \vec{\mu})}^{\pi_1}(T) &=  E \left[ \sum_{t = 0}^{T - 1} \sum_{i \in [N] - i^*} Q^{\pi_1}(t) \mathbf{1} \left\{ u(t) = i \right\} + \sum_{t = 0}^{T-1} Q^{\pi_1}(t) \mathbf{1} \left\{ u(t) = i^* \right\} - \sum_{t = 0}^{T - 1} Q^*(t) \right] \nonumber \\
&\leq E \left[ \sum_{t = 0}^{T - 1} \sum_{i \in [N] - i^*} Q^{\pi_1}(t) \mathbf{1} \left\{ u(t) = i \right\} \right]. \nonumber
\end{align*}

\section{Proof of Lemma~\ref{lemma2}}
\label{lemma2_appendix}

Consider a busy period scheduled to server $i$.  The length of the busy period and integrated backlog over the busy period are random variables $X_i$ and $Z_i$, respectively.  Furthermore,
\begin{equation*}
Z_i(\omega) \leq X_i^2(\omega).
\end{equation*}
We proceed to bound $E\left[X_i^2\right]$.

To this end, define random variable $Y_i$ to be the difference between the number of arrivals and offered service from server $i$ in a time slot.  The variable has probability mass function
\begin{equation*}
P(Y_i = y) = \left\{
\begin{array}{ll}
	\mu_i (1 - \lambda), &y = -1\\
	\mu_i\lambda + (1 - \mu_i)(1 - \lambda), &y = 0\\
	\lambda (1 - \mu_i), &y = 1\\
	0, &\mbox{ otherwise}
\end{array}
\right. .
\end{equation*}
The mean of $Y_i$ is denoted $\overline{Y}_i$ and note that by Assumption~\ref{assumption1} it is negative.  Let $Y_i^k$ for $k = 1, 2, \dots$ be \emph{i.i.d.} random variables with the distribution of $Y_i$.

Then, the duration of the busy period (which begins with one packet in queue) can be modeled as the $-1$ hitting time of a random walk defined by $Y_i$. i.e.,
\begin{equation*}
X_i = \min \left\{n : Y^1_i + Y^2_i + \dots + Y^n_i = -1 \right\}.
\end{equation*}
Note that each entry $Y^k_i$ accounts for the difference in the number of arrivals and amount of service at a time step of the busy period.  Thus, the $-1$ hitting time is the first time slot when the service dominates the arrivals and the queue empties.

Given the above,
\begin{equation}
\label{lower_bounding_X_i}
P(X_i > n) \leq P\left(\sum_{m = 1}^n Y^m_i > -1\right) = P\left( \frac{1}{n} \sum_{m = 1}^n Y^m_i - \overline{Y}_i > -\frac{1}{n} - \overline{Y}_i\right)
\end{equation}
where the inequality follows from the fact that the $-1$ hitting time must have occurred no later than time step $n$ if $\sum_{m = 1}^n Y^m_i$ is found to be less than or equal to $-1$. The equality follows from simple algebra.

Then, for any chosen $\delta \in \left(0, -\overline{Y}_i\right)$, we have for $n > n_0 \triangleq \left \lceil \frac{1}{-\overline{Y}_i - \delta} \right \rceil$,
\begin{equation}
\label{continue_lower_bounding_X_i}
\eqref{lower_bounding_X_i} \leq e^{-\frac{1}{2} n \left(-\frac{1}{n} - \overline{Y}_i\right)^2} \leq e^{-\frac{1}{2} n \delta^2}
\end{equation}
where the first inequality follows from Hoeffding's inequality and the second inequality follows from $n > n_0$.  Thus,
\begin{equation*}
E\left[X_i^2\right] = \sum_{n=1}^\infty n^2 P(X_i = n) \leq \sum_{n=1}^\infty n^2 P(X_i > n-1) \leq \sum_{n = 1}^{n_0} n^2 + \sum_{n = n_0 + 1}^\infty n^2 e^{-\frac{1}{2} (n-1) \delta^2} < \infty.
\end{equation*}
In the above, the second inequality follows from bounding $P(X_i > n - 1)$ with $1$ for the the first $n_0$ terms and \eqref{continue_lower_bounding_X_i} for all terms greater than $n_0$.  Likewise, the last inequality follows from $\sum_{n = 0}^\infty n^2 e^{-\frac{1}{2} n \delta^2} < \infty$.  This establishes the result.

\section{Proof of Lemma~\ref{lemma3}}
\label{lemma3_appendix}

The proof follows from the fact that, with probability one, there can be no greater than $T$ busy periods in the time interval $0$ to $T-1$.  Thus, the cost accrued over the busy periods occurring before time $T$, must be less than the cost accrued over the first $T$ busy periods.  Concretely, denoting the start time and finish time of busy period $p$ with $s_p$ and $f_p$, we have that
\begin{equation}
\label{bounded_by_number_of_busy_periods}
E \left[ \sum_{t = 0}^{T - 1} Q^{\pi_1}(t) \mathbf{1} \left\{ u(t) = i \right\} \right] \leq E\left[ \sum_{p = 1}^{T} \mathbf{1} \left\{  i \mbox{ scheduled for busy period } p \right\} \sum_{\tau = s_p}^{f_p} Q^{\pi_1}(\tau) \right].
\end{equation}
Thus, using the definition of $Z_i$ and $S^{\pi_1}_i(T)$, 
\begin{align*}
\eqref{bounded_by_number_of_busy_periods} &= \sum_{p = 1}^{T} P \left(  i \mbox{ scheduled for busy period } p \right) E \left[ \sum_{\tau = s_p}^{f_p} Q^{\pi_1}(\tau) \bigg \rvert i \mbox{ scheduled for busy period } p  \right] \nonumber \\
&= \overline{Z}_i \sum_{p = 1}^{T} P \left(  i \mbox{ scheduled for busy period } p \right) \nonumber \\
&=  \overline{Z}_i E \left[ S^{\pi_1}_i(T) \right].
\end{align*}

\section{Proof of Lemma~\ref{lemma4}}
\label{lemma4_appendix}

Define $\gamma \triangleq \frac{1}{2} \left(\mu_{i^*} - \mu_i \right)$.  Note that $\gamma$ is the half way distance between $\mu_{i^*}$ and $\mu_{i}$.  Furthermore, define $\widehat{\mu}_i(p)$ to be the sample mean that policy $\pi_1$ has for server $i$ at the start of busy period $p$ and $T_i(p)$ the number of times the policy explored $i$ during an empty period before busy period $p$ (i.e., the number of times the policy has updated $\widehat{\mu}_i(p)$).  Recall $\widehat{\mu}_i(p) = 0$ if we have not yet sampled $i$.  Then,
\begin{align*}
\left\{i \mbox{ scheduled for busy period } p \right\} & \subseteq
\left\{\widehat{\mu}_{i^*}(p) \leq \widehat{\mu}_{i}(p)\right\} \nonumber \\
&\subseteq
\left\{ \left\{ \widehat{\mu}_i(p) - \mu_i \geq \gamma \right\} \cup \left\{ \widehat{\mu}_{i^*}(p) - \mu_{i^*} \leq -\gamma \right\} \right\}.
\end{align*}
The above states that in order to schedule $i$ during $p$, $i$ must have a sample mean no less than $i^*$, and this only occurs if one of the sample means deviates from its true mean by a distance $\gamma$.

Now, for a sample mean to deviate by distance $\gamma$, the policy must either have not observed $i$ very often during exploration or it must have observed atypical observations during exploration.  Thus, we further break up the above events:
\begin{equation*}
\left\{ \widehat{\mu}_i(p) - \mu_i \geq \gamma \right\} \subseteq \left\{ \left\{ T_i(p) \leq \frac{p}{2N} \right\} \cup \left\{ \left\{ T_i(p) > \frac{p}{2N} \right\} \cap \left\{ \widehat{\mu}_i(p) - \mu_i \geq \gamma \right\} \right\} \right\}
\end{equation*}
and
\begin{equation*}
\left\{ \widehat{\mu}_{i^*}(p) - \mu_{i^*} \leq -\gamma \right\} \subseteq \left\{ \left\{ T_{i^*}(p) \leq \frac{p}{2N} \right\} \cup \left\{ \left\{ T_{i^*}(p) > \frac{p}{2N} \right\} \cap \left\{ \widehat{\mu}_{i^*}(p) - \mu_{i^*} \leq -\gamma \right\} \right\} \right\}.
\end{equation*}

Note that $T_i(p)$ and $T_{i^*}(p)$ are sums of Bernoulli random variables that take the value $1$ when $i$ is observed during empty period $p$ and that $\frac{1}{p} E\left[T_i(p)\right] = \frac{1}{p} E\left[T_{i^*}(p)\right] = \frac{1}{N}$. Using Hoeffding's inequality,
\begin{equation*}
P\left(T_i(p) \leq \frac{p}{2N}\right) = P\left(\frac{1}{p} T_i(p) - \frac{1}{N} \leq -\frac{1}{2N} \right) \leq e^{-2 p \left( \frac{1}{2N} \right)^2}
\end{equation*}
with a similar bound holding for $P\left(T_{i^*}(p) \leq \frac{p}{2N}\right)$.

Now, the number of times we have explored a server during the empty periods, is independent of the observations that were made.  Therefore, we can use Hoeffding's to obtain
\begin{equation*}
P\left(\widehat{\mu}_{i^*}(p) - \mu_{i^*} \leq -\gamma \left \lvert T_i(p) > \frac{p}{2N} \right. \right) \leq e^{-2 \frac{p}{2N} \gamma^2 }
\end{equation*}
and
\begin{equation*}
P\left(\widehat{\mu}_{i}(p) - \mu_{i} \geq \gamma \left \lvert T_i(p) > \frac{p}{2N} \right. \right) \leq e^{-2 \frac{p}{2N} \gamma^2 }.
\end{equation*}
Trivially, $P\left(T_i(p) > \frac{p}{2N}\right)$ and $P\left(T_{i^*}(p) > \frac{p}{2N}\right)$ are bounded above by $1$.

Then, applying the union bound and the above concentration inequalities, we obtain
\begin{align*}
E\left[S^{\pi_1}_i(T)\right] &= \sum_{p = 1}^{T} P\left( i \mbox{ scheduled for busy period } p \right) \nonumber \\
&\leq \sum_{p = 1}^{T} \left( 2 e^{-2 p \left( \frac{1}{2N} \right)^2} + 2 e^{-2 \frac{p}{2N} \gamma^2 } \right) \nonumber \\
& = O(1). \nonumber
\end{align*}

\section{Proof of Lemma~\ref{lemma5}}
\label{lemma5_appendix}
Recall that $C(p)$ is an indicator that must equal $1$ if we schedule a server not equal to $i^*$ at any point during period $p$, and $p(t)$ denotes the period (empty or busy) that time slot $t$ resides in.

The proof of Lemma~\ref{lemma5} closely follows the proof of Lemma~\ref{lemma1}.  Specifically, for any outcome $\omega$ consider a time slot $t$ such that $C(p(t, \omega), \omega) = 0$ and $Q^{\pi_2}(t, \omega) > 0$.\footnote{We use the notation $C(p, \omega)$ to denote the value of the indicator $C(p)$ for given outcome $\omega$.}  Then, over this sample path, the server $i^*$ is scheduled over the entire busy period in which time slot $t$ resides.  By an argument similar to Lemma~\ref{lemma1}, this implies that $Q^{\pi_2}(t, \omega) \leq Q^*(t, \omega)$.  Therefore, summing over $t$
\begin{equation*}
\sum_{t = 0}^{T-1} Q^{\pi_2}(t, \omega) \mathbf{1} \left\{ C(p(t, \omega), \omega) = 0 \right\} \leq \sum_{t = 0}^{T - 1} Q^*(t, \omega).
\end{equation*}
Taking expectation,
\begin{equation*}
E\left[\sum_{t = 0}^{T-1} Q^{\pi_2}(t) \mathbf{1} \left\{ C(p(t)) = 0 \right\}\right] \leq E\left[\sum_{t = 0}^{T - 1} Q^*(t) \right].
\end{equation*}
Then using the definition of queue length regret,
\begin{align*}
R_{(\lambda, \vec{\mu})}^{\pi_2}(T) &=  E \left[ \sum_{t = 0}^{T - 1} Q^{\pi_2}(t) \mathbf{1} \left\{ C(p(t)) = 1 \right\} + \sum_{t = 0}^{T-1} Q^{\pi_2}(t) \mathbf{1} \left\{ C(p(t)) = 0 \right\} - \sum_{t = 0}^{T - 1} Q^*(t) \right] \nonumber \\
&\leq E \left[ \sum_{t = 0}^{T - 1} Q^{\pi_2}(t) \mathbf{1} \left\{ C(p(t)) = 1 \right\} \right] \nonumber \\
&= E \left[ \sum_{t = 0}^{T-1} Q^{\pi_2}(t) C(p(t)) \right]. \nonumber
\end{align*}

\section{Proof of Lemma~\ref{lemma6}}
\label{lemma6_appendix}

Define $C_{i^*}(p)$ be an indicator random variable that equals $1$ if we scheduled $i^*$ for the start of busy period $p$ and hit the time-out threshold (i.e., the queue did not empty in the first $p$ time slots).  Otherwise, let $C_{i^*}(p)$ be $0$.  Then,
\begin{equation*}
\left\{C(p) = 1\right\} = \left\{ \left\{ C_{i^*}(p) = 1 \right\} \cup \left\{i^* \mbox{ not scheduled for start of busy period } p \right\} \right\}.
\end{equation*}

Define $\widehat{\mu}_i(p)$ to be the sample mean that policy $\pi_2$ has for server $i$ at the start of busy period $p$. Now, if we did not schedule $i^*$ for the start of busy period $p$, then there must exist an $i \neq i^*$ such that $\widehat{\mu}_{i^*}(p) \leq \widehat{\mu}_{i}(p)$.  Thus, using the union bound,
\begin{equation}
\label{event1_bound}
P\left( C(p) = 1 \right) \leq P(C_{i^*}(p) = 1) + \sum_{i = [N] - i^*} P\left( \widehat{\mu}_{i^*}(p) \leq \widehat{\mu}_{i}(p) \right).
\end{equation}

Define $\gamma_i \triangleq \frac{1}{2} \left(\mu_{i^*} - \mu_i \right)$ and $\gamma \triangleq \min_{i \in [N] - i^*} \gamma_i$.
Then, by analysis similar to proof of Lemma~\ref{lemma4},
\begin{equation}
\label{event2_bound}
 P\left( \widehat{\mu}_{i^*}(p) \leq \widehat{\mu}_{i}(p) \right) \leq 2 e^{-2 p \left( \frac{1}{2N} \right)^2} + 2 e^{-2 \frac{p}{2N} \gamma^2 }.
\end{equation}

All that is left is to characterize $P(C_{i^*}(p) = 1)$.  To do so, we will analyze the probability that the time-out threshold $p$ is reached given $i^*$ is chosen to be scheduled at the start of busy period $p$.  For an arbitrary time slot, the difference in the number of arrivals and available service from $i^*$ is given by a random variable $Y_{i^*}$ that has probability mass function
\begin{equation*}
P(Y_{i^*} = y) = \left\{
\begin{array}{ll}
	\mu_{i^*} (1 - \lambda), &y = -1\\
	\mu_{i^*}\lambda + (1 - \mu_{i^*})(1 - \lambda), &y = 0\\
	\lambda (1 - \mu_{i^*}), &y = 1\\
	0, &\mbox{ otherwise}
\end{array}
\right. .
\end{equation*}
Then, let random variable $Y^k_{i^*}$ be the difference in the number of arrivals and available service from $i^*$ in the $k^{th}$ time slot since the start of the busy period.  Note that the random variables are \emph{i.i.d.} with the distribution of $Y_{i^*}$ above.

Given that $i^*$ is scheduled at the start of busy period $p$, we will analyze the accumulated difference between the number of arrivals and service process (since the start of the busy period) as a random walk $Y^1_{i^*} + Y^2_{i^*} + \dots$.  The busy period begins with one packet in queue.  Therefore, if the random walk hits $-1$ in $p$ time slots, the queue empties and the period ends before reaching the time-out threshold.  Note that by Assumption~\ref{assumption2}, the mean $\overline{Y}_{i^*} < 0$ and the walk has negative drift.
Then,
\begin{align}
P(C_{i^*}(p) = 1) &= P\left(\min_{n \leq p} \left\{Y_{i^*}^1 + Y_{i^*}^2 + \dots + Y_{i^*}^n \right\} > - 1 \right) \nonumber \\
& \leq P\left( \sum_{m = 1}^{p} Y_{i^*}^m > -1 \right) \nonumber \\  
&= P\left( \frac{1}{p} \sum_{m = 1}^{p} Y_{i^*}^m - \overline{Y}_{i^*} > -\frac{1}{p} - \overline{Y}_{i^*} \right). \label{prob_hit_timeout}
\end{align}
Now, for any choice of $\delta \in \left(0, -\overline{Y}_{i^*}\right)$, we have for all $p \geq p_0 \triangleq \left \lceil \frac{1}{-\overline{Y}_{i^*} - \delta} \right \rceil$,
\begin{equation}
\label{event3_bound}
\eqref{prob_hit_timeout} \leq e^{-\frac{1}{2} p \left( -\frac{1}{p} - \overline{Y}_{i^*} \right)^2 } \leq e^{-\frac{1}{2}p\delta^2}
\end{equation}
where the first inequality follows from Hoeffding's inequality and the second inequality from $p \geq p_0$.

Plugging in \eqref{event2_bound} and \eqref{event3_bound} into \eqref{event1_bound}, we see that for all $p \geq p_0$,
\begin{equation*}
P\left(C(p) = 1\right) \leq e^{-\frac{1}{2} p \delta^2} + (N-1)\left( 2 e^{-2 p \left( \frac{1}{2N} \right)^2} + 2 e^{-2 \frac{p}{2N} \gamma^2 } \right).
\end{equation*}
Then, we can clearly choose values of $M_0$ and $\chi$, that are independent of $p$, to meet the lemma's condition.

\section{Proof of Lemma~\ref{lemma7}}
\label{lemma7_appendix}

We define $U(p)$ to be the amount of time that we schedule the randomized policy over busy period $p$. Then, an upper bound on the duration of the busy period is $p + U(p)$.  Now, in an argument similar to Appendix~\ref{lemma2_appendix}, we can bound the integrated queue backlog over the busy period using
\begin{equation}
\label{upper_bound_on_z}
Z^{\pi_2}(p) \leq (p + U(p))^2.
\end{equation}
We therefore need a bound on the $P(U(p) > n \rvert C(p) = 1)$ for $n = 1, 2, \dots$.

To this end, we analyze the difference between the number of arrivals and offered service (provided by the randomized policy) to the queue that would occur after threshold-time $p$.  Let the average offered service of the randomized policy be $\mu_\alpha \triangleq \sum_{i \in [N]} \alpha_i \mu_i$.  Define random variable $Y_\alpha$ to be the difference between the number of arrivals and service offered by the randomized policy in a given time slot.  This random variable has probability mass function
\begin{equation*}
P(Y_{\alpha} = y) = \left\{
\begin{array}{ll}
	\mu_{\alpha} (1 - \lambda), &y = -1\\
	\mu_{\alpha}\lambda + (1 - \mu_{\alpha})(1 - \lambda), &y = 0\\
	\lambda (1 - \mu_{\alpha}), &y = 1\\
	0, &\mbox{ otherwise}
\end{array}
\right. .
\end{equation*}
Let $Y^1_\alpha, Y^2_\alpha, \dots$ be \emph{i.i.d.} random variables with the distribution of $Y_\alpha$.  Each $Y^k_\alpha$ is the difference in the number of arrivals and offered service in time slot $k$ following the time-out threshold.
By Assumption~\ref{assumption2}, the mean $\overline{Y}_\alpha < 0$ and the random walk $Y^1_\alpha + Y^2_\alpha + \dots$ has negative drift.

Now, given that we hit the time-out threshold, at most $p + 1$ packets are in the queue when we start scheduling the known randomized policy. 
Then, we see that the length of time that we could have possibly scheduled the randomized policy is bounded by: $\{U(p) > n\} \subseteq \{ Y^1_\alpha + Y^2_\alpha + \dots Y^m_\alpha > -(p + 1), \forall m = 1, 2, \dots, n \}$ (i.e., the random walk has yet to hit $-(p + 1)$).  Note that this is a pessimistic bound, since there may be less than $p + 1$ packets in the queue at the start of scheduling the randomized policy.  Then, we can obtain for $n \geq 1$,
\begin{align}
P\left(U(p) > n \rvert C(p) = 1 \right) &\leq P\left( Y^1_\alpha + Y^2_\alpha + \dots Y^m_\alpha > -(p + 1), \forall m = 1, 2, \dots, n \right) \nonumber \\
& \leq P\left( \sum_{m = 1}^n Y_\alpha^m > -(p + 1) \right) \nonumber \\
& \leq P\left( \frac{1}{n} \sum_{m=1}^n Y_\alpha^m - \overline{Y}_\alpha > -\frac{p + 1}{n} - \overline{Y}_\alpha \right) \label{bound_on_recovery}
\end{align}
where the second inequality follows from the simple fact that if $\sum_{m = 1}^n Y_\alpha^m \leq -(p + 1)$ we have clearly already hit $-(p+1)$, and the third inequality follows from simple algebra.

Now, for any choice of $\delta \in \left(0,  -\overline{Y}_\alpha \right)$, for $n \geq n_0(p) \triangleq \left \lceil \frac{p + 1}{-\overline{Y}_\alpha - \delta} \right \rceil$,
\begin{equation*}
\eqref{bound_on_recovery} \leq e^{-\frac{1}{2} n \left( -\frac{p + 1}{n} - \overline{Y}_\alpha \right)^2} \leq e^{-\frac{1}{2} n \delta^2}
\end{equation*}
where the first inequality follows from Hoeffding's inequality and the second from $n \geq n_0(p)$.  Importantly, note that $n_0(p)$ is a function of $p$.

Putting this together, starting at \eqref{upper_bound_on_z}, we see that
\begin{align*}
E\left[\left. Z^{\pi_2}(p) \right \rvert C(p) = 1 \right] &\leq \sum_{n=0}^\infty (p + n)^2 P\left(U(p) = n \rvert C(p) = 1 \right) \nonumber \\
& = \sum_{n=0}^{n_0(p)} (p + n)^2 P\left(U(p) = n \rvert C(p) = 1 \right) + \sum_{n=n_0(p) + 1}^\infty (p + n)^2 P\left(U(p) = n \rvert C(p) = 1 \right) \nonumber \\
& \leq  \sum_{n=0}^{n_0(p)} (p + n)^2 P\left(U(p) = n \rvert C(p) = 1 \right) + \sum_{n=n_0(p) + 1}^\infty (p + n)^2 P\left(U(p) > n - 1 \rvert C(p) = 1 \right) \nonumber \\
&\leq \left(p  + \left\lceil\frac{p + 1}{-\overline{Y}_\alpha - \delta}\right\rceil \right)^2 + \sum_{n=0}^\infty \left(p + n \right)^2 e^{-\frac{1}{2} n \delta^2} \nonumber \\
& 
\end{align*}
Then, since $\sum_{n=0}^\infty n e^{-\frac{1}{2}n \delta^2} < \infty$ and $\sum_{n=0}^\infty n^2 e^{-\frac{1}{2}n \delta^2} < \infty$, 
it is easy to see that we can choose values of $M_1$ and $\beta_1$ such that the lemma holds.

\section{Proof of Lemma~\ref{lemma8}}
\label{lemma8_appendix}

Recall that $Z^{\pi_3}(p)$ is the total queue backlog summed over the interval of busy period $p$.
We begin by defining $U(p)$ to be the amount of time that the learning method of Fig.~\ref{policy_learning} is used during busy period $p$.  Then, in an argument similar to the proof of Lemma~\ref{lemma7},
\begin{equation}
\label{bound_on_z_s}
Z^{\pi_3}(p) \leq \left(p + U(p)\right)^2.
\end{equation}
We seek to bound $P(U(p) > n \rvert C(p) = 1)$ for $n = 1, 2, \dots$.

In the following, we will use $\widehat{\mu}^p_i(n)$ to denote the learning algorithm's estimate of server $i$ at time slot $n$ (see Fig.~\ref{policy_learning}) and $T^p_i(n)$ the number of times the algorithm has observed server $i$ during an exploration (i.e., the number of times $\widehat{\mu}^p_i(n)$ has been updated).  Recall that $n$ takes value $0$ when the timeout threshold $p$ is crossed. (i.e., Variable $n$ counts the number of time slots since the threshold was crossed.)  Likewise, we have that the number of explorations made by time slot $n$ is bounded by
\begin{equation}
\label{bound_on_vn}
r n^\epsilon - b_1 \leq V(n) \leq r n^\epsilon + b_2
\end{equation}
for positive constants $b_1, b_2, r$, and $\epsilon \in (0, 1)$.
Finally, we define a constant $\sigma \in (0,1)$ such that $\frac{\lambda}{\mu_{i^*}} < 1 - \sigma$, which by Assumption~\ref{assumption3} must exist.

In order to analyze the event $\{U(p) > n\}$, we consider whether running the learning algorithm of Fig.~\ref{policy_learning} over $n$ time slots would have caused the queue to empty at any time up to or before $n$.  Recall that at most $p + 1$ packets can be in the queue when the learning method is initiated.
We will then break up the event $\{U(p) > n\}$ into two parts.  First, the event that $\widehat{\mu}^p_{i^*}$ was not the maximum estimate for any time slot over the interval $(\sigma n, n - 1)$.  Second, the event that $\widehat{\mu}^p_{i^*}$ was the maximum estimate over the interval $(\sigma n, n - 1)$ (and, thus, at all non-exploration time slots $i^*$ was scheduled), but the arrival process plus the initial queue backlog (which can be no greater than $p + 1$) has dominated the service process up until time $n$.  Concretely, for $n \geq 1$,
\begin{multline}
\label{bound_on_u_p}
\left\{U(p) > n\right\} \subseteq \Bigg\{
\Bigg\{\exists m \in \left(\sigma n, n - 1\right), \exists i \in [N] - i^* : \widehat{\mu}^p_{i^*}(m) \leq \widehat{\mu}^p_{i}(m) \Bigg\} \\ \cup \Bigg\{ \Bigg\{\widehat{\mu}^p_{i^*}(m) > \widehat{\mu}^p_{i}(m), \forall m \in \left(\sigma n, n - 1\right), \forall i \in [N] - i^* \Bigg\} \cap \left\{\sum_{m = 0}^{n-1} A(m) + p + 1 > \sum_{m = 0}^{n-1} D(m) \right\}\Bigg\} \Bigg\}.
\end{multline}

We now use the following lemmas.  Their proofs can be found at the end of this appendix.
\begin{lemma}
\label{lemma_8_1}
For positive constants $c_1$ and $c_2$ that are independent of $n$ and $p$,
\begin{equation*}
P\left( \exists m \in \left(\sigma n, n - 1\right), \exists i \in [N] - i^* : \widehat{\mu}^p_{i^*}(m) \leq \widehat{\mu}^p_{i}(m) \right) \leq c_1 n e^{-c_2 n^\epsilon}.
\end{equation*}
\end{lemma}

\begin{lemma}
\label{lemma_8_2}
For $n \geq c_4 + c_5 p$,
\begin{equation*}
P\Bigg( \sum_{m = 0}^{n-1} A(m) + p + 1 > \sum_{m = 0}^{n - 1} D(m) \Bigg\lvert \widehat{\mu}^p_{i^*}(m) > \widehat{\mu}^p_{i}(m), \forall m \in \left(\sigma n, n - 1\right), \forall i \in [N] - i^* \Bigg) \leq 2e^{-c_3 n}
\end{equation*}
where $c_3$, $c_4$, and $c_5$ are positive constants that are independent of $n$ and $p$.
\end{lemma}

We now use Lemmas~\ref{lemma_8_1}~and~\ref{lemma_8_2} to construct a bound on $E\left[Z^{\pi_3}(p) \Big \rvert C(p) = 1 \right]$.
\begin{align}
E\left[Z^{\pi_3}(p) \Big \rvert C(p) = 1 \right] & \leq \sum_{n = 0}^\infty (p + n)^2 P(U(p) = n \rvert C(p) = 1 ) \nonumber \\
& = \sum_{n = 0}^{\lceil c_4 + c_5 p \rceil} (p + n)^2 P\left( U(p) = n \rvert C(p) = 1 \right) + \sum_{n = \lceil c_4 + c_5 p \rceil + 1}^\infty (p + n)^2 P\left( U(p) = n \rvert C(p) = 1 \right) \nonumber \\
& \leq \left( p + \lceil c_4 + c_5 p \rceil \right)^2 + \sum_{n = \lceil c_4 + c_5 p \rceil + 1}^\infty (p + n)^2 P\left( U(p) > n - 1 \rvert C(p) = 1 \right) \nonumber \\
& \leq \left( p + \lceil c_4 + c_5 p \rceil \right)^2 + \sum_{n = 0}^\infty \left( p + n \right)^2 \left( (c_1 n) e^{-c_2 n^\epsilon} + 2e^{-c_3 n} \right) \label{almost_there}
\end{align}
where the first inequality follows from \eqref{bound_on_z_s} and the last inequality follows from applying Lemmas~\ref{lemma_8_1}~and~\ref{lemma_8_2} and the union bound to \eqref{bound_on_u_p}.
Note that for any non-negative integer $x$, real number $y > 0$, and $z \in (0,1]$,
\begin{equation*}
\sum_{n = 0}^\infty n^x e^{-y n^z} < \infty.
\end{equation*}
Then one can see that \eqref{almost_there} can be bounded by a quadratic function of $p$ giving the result.  Below we provide the proofs of Lemmas~\ref{lemma_8_1}~and~\ref{lemma_8_2}.

\begin{IEEEproof}[Proof of Lemma~\ref{lemma_8_1}]

We proceed to bound the probability of 
\begin{equation*}
\Bigg\{ \exists m \in \left(\sigma n, n - 1\right), \exists i \in [N] - i^* : \widehat{\mu}^p_{i^*}(m) \leq \widehat{\mu}^p_{i}(m) \Bigg\}.
\end{equation*}

To this end, we can apply the following bounds

\begin{align}
P\left( \exists m \in \left(\sigma n, n - 1\right), \exists i \in [N] - i^* : \widehat{\mu}^p_{i^*}(m) \leq \widehat{\mu}^p_{i}(m) \right) & \leq \sum_{m = \lceil \sigma n \rceil}^{n - 1} P\left( \exists i \in [N] - i^* : \widehat{\mu}^p_{i^*}(m) \leq \widehat{\mu}^p_{i}(m) \right) \nonumber \\
& \leq \sum_{m = \lceil \sigma n \rceil}^{n - 1} \sum_{i \in [N] - i^*} P \left( \widehat{\mu}^p_{i^*}(m) \leq \widehat{\mu}^p_{i}(m) \right). \label{not_scheduling_best}
\end{align}
We proceed via an argument similar to the proof of Lemma~\ref{lemma4}.  Consider an $m \geq \lceil \sigma n \rceil$.   Define 
\begin{equation*}
\gamma \triangleq \min_{i \in [N] -i^*} \frac{1}{2} \left( \mu_{i^*} - \mu_i \right).
\end{equation*}
Then, as was argued in Lemma~\ref{lemma4},
\begin{equation*}
\left\{\widehat{\mu}^p_{i^*}(m) \leq \widehat{\mu}^p_{i}(m)\right\} \subseteq \left\{ \left\{ \widehat{\mu}^p_i(m) - \mu_i \geq \gamma \right\} \cup \left\{ \widehat{\mu}^p_{i^*}(m) - \mu_{i^*} \leq -\gamma \right\} \right\}.
\end{equation*}
Recall that by time $m$, $V(m)$ dedicated explore time slots have occurred.  We proceed to bound the two events on the right-hand side of the above. As in Lemma~\ref{lemma4}, we break up these events into two cases; the event the method under observes a server $i$ and the event the method observes atypical observations of the server. Concretely,
\begin{equation*}
\left\{ \widehat{\mu}^p_i(m) - \mu_i \geq \gamma \right\} \subseteq \left\{ \left\{ T^p_i(m) \leq \frac{V(m)}{2N} \right\} \cup \left\{ \left\{ T^p_i(m) > \frac{V(m)}{2N} \right\} \cap \left\{ \widehat{\mu}^p_i(m) - \mu_i \geq \gamma \right\} \right\} \right\}
\end{equation*}
and
\begin{equation*}
\left\{ \widehat{\mu}^p_{i^*}(m) - \mu_{i^*} \leq -\gamma \right\} \subseteq \left\{ \left\{ T^p_{i^*}(m) \leq \frac{V(m)}{2N} \right\} \cup \left\{ \left\{ T^p_{i^*}(m) > \frac{V(m)}{2N} \right\} \cap \left\{ \widehat{\mu}^p_{i^*}(m) - \mu_{i^*} \leq -\gamma \right\} \right\} \right\}.
\end{equation*}

The probability of under observing $i$ is then given by\footnote{Note that this bound holds even for $V(m) = 0$ since then $e^{-2 (V(m)) \left( \frac{1}{2N} \right)^2}$ trivially equals $1$.}
\begin{equation*}
P\left( T_i^p(m) \leq \frac{V(m)}{2N} \right) = P\left( \frac{1}{V(m)} T_i^p(m) - \frac{1}{N} \leq - \frac{1}{2N} \right) \leq e^{-2 (V(m)) \left( \frac{1}{2N} \right)^2} \leq e^{-2 (r m^\epsilon - b_1)\left( \frac{1}{2N} \right)^2} \leq e^{-2 (r (\sigma n)^\epsilon - b_1)\left( \frac{1}{2N} \right)^2}
\end{equation*}
where the first inequality follows from Hoeffding's inequality, the second inequality follows from plugging in the lower bound of \eqref{bound_on_vn}, and the third inequality follows from $m \geq \lceil \sigma n \rceil$. A similar expression holds for $T^p_{i^*}(m)$.  We next proceed to bound the probability that the server sees atypical observations.
\begin{equation*}
P\left( \widehat{\mu}^p_i(m) - \mu^p_i \geq \gamma  \bigg\rvert T^p_i(m) > \frac{V(m)}{2 N} \right) \leq e^{-2 \frac{V(m)}{2N} \gamma^2 } \leq e^{-2 \frac{r m^\epsilon - b_1}{2N} \gamma^2 } \leq e^{-2 \frac{r (\sigma n)^\epsilon - b_1}{2N} \gamma^2 }
\end{equation*}
and
\begin{equation*}
P\left( \widehat{\mu}^p_{i^*}(m) - \mu^p_{i^*} \leq -\gamma  \bigg\rvert T^p_{i^*}(m) > \frac{V(m)}{2 N} \right) \leq e^{-2 \frac{V(m)}{2N} \gamma^2 } \leq e^{-2 \frac{r m^\epsilon - b_1}{2N} \gamma^2 } \leq e^{-2 \frac{r (\sigma n)^\epsilon - b_1}{2N} \gamma^2 }.
\end{equation*}
Thus, putting this altogether starting from \eqref{not_scheduling_best}, we see that
\begin{align}
&P\left( \exists m \in \left(\sigma n, n - 1\right), \exists i \in [N] - i^* : \widehat{\mu}^p_{i^*}(m)  \leq \widehat{\mu}^p_{i}(m) \right) \nonumber \\
&\qquad \qquad \leq \sum_{m = \lceil \sigma n \rceil}^{n - 1} \sum_{i \in [N] - i^*} \left( 2 e^{-2 (r (\sigma n)^\epsilon - b_1)\left( \frac{1}{2N} \right)^2} + 2 e^{-2 \frac{r (\sigma n)^\epsilon - b_1}{2N} \gamma^2 }\right) \nonumber \\
& \qquad \qquad \leq (n - \sigma n)(N - 1) \left( 2 e^{-2 (r (\sigma n)^\epsilon - b_1)\left( \frac{1}{2N} \right)^2} + 2 e^{-2 \frac{r (\sigma n)^\epsilon - b_1}{2N} \gamma^2 }\right) \nonumber \\
& \qquad \qquad \leq c_1 n e^{-c_2 n^\epsilon} \nonumber
\end{align}
where $c_1$ and $c_2$ are positive constants that are independent of $n$ and $p$.  This gives the lemma.
\end{IEEEproof}

\begin{IEEEproof}[Proof of Lemma~\ref{lemma_8_2}]

We wish to bound
\begin{equation*}
P\left( \sum_{m = 0}^{n-1} A(m) + p + 1 > \sum_{m = 0}^{n - 1} D(m) \bigg\lvert \widehat{\mu}^p_{i^*}(m) > \widehat{\mu}^p_{i}(m), \forall m \in \left(\sigma n, n - 1\right), \forall i \in [N] - i^* \right).
\end{equation*}

We begin by defining $\tau(n)$ to be the set of time slots that are not dedicated exploration time slots in the interval $(\sigma n, n-1)$.  Let $D^*(m)$ be the service available from server $i^*$ at time slot $m$.  Then we can bound,
\begin{align}
&P\left( \sum_{m = 0}^{n-1} A(m) + p + 1 > \sum_{m = 0}^{n - 1} D(m) \bigg\lvert \widehat{\mu}^p_{i^*}(m) > \widehat{\mu}^p_{i}(m), \forall m \in \left(\sigma n, n - 1\right), \forall i \in [N] - i^* \right) \nonumber \\
& \qquad \leq P\left( \sum_{m = 0}^{n-1} A(m) + p + 1 > \sum_{m \in \tau(n) } D^*(m)  \right) \nonumber \\
&\qquad \leq P\left( \sum_{m = 0}^{n-1} A(m) + p + 1 \geq \frac{1}{2} \left( (1 - \sigma) \mu_{i^*} + \lambda \right) n \right) + P\left( \sum_{m \in \tau(n)} D^*(m) \leq \frac{1}{2} \left( (1 - \sigma) \mu_{i^*} + \lambda \right) n \right) \label{arrivals_dominate_departures0}
\end{align}
where the last inequality follows by defining a threshold $\frac{1}{2} \left( (1 - \sigma) \mu_{i^*} + \lambda \right) n$ and noting that $\sum_{m = 0}^{n-1} A(m) + p + 1 \leq \sum_{m \in \tau(n) } D^*(m)$ if  $\sum_{m = 0}^{n-1} A(m) + p + 1$ is less than the threshold and $\sum_{m \in \tau(n)} D^*(m)$ is greater than the threshold.

We now turn to analyzing how the cardinality of $\tau(n)$ scales with $n$.  We begin by noting that for $n - 1 < \lceil \sigma n \rceil$, $\lvert \tau(n) \rvert = 0$.  Thus, the set is empty for relatively small $n$.  For $n$ large enough such that $n - 1 \geq \lceil \sigma n \rceil$, we have
\begin{equation}
\label{cardinality_of_tau}
(n - 1) - \lceil \sigma n \rceil + 1 - (r n^\epsilon + b_2) \leq \lvert \tau(n) \rvert \leq (n - 1) - \lceil \sigma n \rceil + 1
\end{equation}
since there are $(n - 1) - \lceil \sigma n \rceil + 1$ time slots in the interval $(\sigma n, n - 1)$, at most $V(n)$ of them are dedicated exploration time slots, and we can use \eqref{sublinear_exploration} to upper bound $V(n)$.  Using \eqref{cardinality_of_tau}, we see that\footnote{To be concrete, this is technically true for the subsequence of $\frac{n}{\lvert \tau(n) \rvert}$ for which $\lvert \tau(n) \rvert > 0$ and for $n > 0$.  However, this technicality will not impact our subsequent analysis.}
\begin{align}
\lim_{n \to \infty} \frac{\lvert \tau(n) \rvert}{n} &= 1 - \sigma \nonumber \\
\lim_{n \to \infty} \frac{n}{\lvert \tau(n) \rvert} &= \frac{1}{1 - \sigma}. \label{limit_n_over_tau}
\end{align}

Given the above, for $n$ large enough such that $\lvert \tau(n) \rvert \geq 1$, we use simple algebra to make the following bound on \eqref{arrivals_dominate_departures0}.
\begin{multline}
\eqref{arrivals_dominate_departures0} \leq P\left( \frac{1}{n} \sum_{m = 0}^{n-1} A(m) - \lambda \geq \frac{1}{2} \left( (1 - \sigma) \mu_{i^*} -  \lambda \right) - \frac{p + 1}{n} \right) \\ + P\left( \frac{1}{\left \lvert \tau(n) \right \rvert} \sum_{m \in \tau(n)} D^*(m) - \mu_{i^*} \leq \frac{1}{2}\left( (1 - \sigma) \mu_{i^*} + \lambda \right) \frac{n}{\lvert \tau(n) \rvert} - \mu_{i^*} \right).
\label{arrivals_dominate_departures1}
\end{multline}
Now, using \eqref{limit_n_over_tau}, we see that
\begin{align*}
\lim_{n \to \infty} \left(\frac{1}{2}\left( (1 - \sigma) \mu_{i^*} + \lambda \right) \frac{n}{\lvert \tau(n) \rvert} - \mu_{i^*}\right) &= -\frac{1}{1 - \sigma} \frac{1}{2} \left( (1 - \sigma) \mu_{i^*} - \lambda \right) \nonumber \\
& < -\frac{1}{2} \left( (1 - \sigma) \mu_{i^*} - \lambda \right) \nonumber
\end{align*}
where the last inequality follows from the fact that $\frac{1}{1 - \sigma} > 1$.  Thus, there exists an $n_0$ that is independent of $p$ such that $\forall n \geq n_0$
\begin{multline}
\label{arrivals_dominate_departures2}
\eqref{arrivals_dominate_departures1} \leq P\left( \frac{1}{n} \sum_{m = 0}^{n-1} A(m) - \lambda \geq \frac{1}{2} \left( (1 - \sigma) \mu_{i^*} -  \lambda \right) - \frac{p + 1}{n} \right) \\ + P\left( \frac{1}{\left \lvert \tau(n) \right \rvert} \sum_{m \in \tau(n)} D^*(m) - \mu_{i^*} \leq  -\frac{1}{2} \left( (1 - \sigma) \mu_{i^*} - \lambda \right)  \right).
\end{multline}
To declutter notation, we let $\delta \triangleq \frac{1}{2} \left( (1 - \sigma) \mu_{i^*} - \lambda \right)$.  Note that by the definition of $\sigma$, $\delta > 0$.

Now, for any $\theta \in (0, \delta)$, if $n$ is large enough such that $n \geq \frac{p + 1}{\delta - \theta}$ we have that
\begin{equation*}
\delta - \frac{p + 1}{n} \geq \theta.
\end{equation*} 
Thus, for $n \geq \max\{n_0, \frac{p + 1}{\delta - \theta} \}$ we can bound \eqref{arrivals_dominate_departures2} as
\begin{align}
\eqref{arrivals_dominate_departures2} &\leq P\left( \frac{1}{n} \sum_{m = 0}^{n-1} A(m) - \lambda \geq \theta \right) + P\left( \frac{1}{\left \lvert \tau(n) \right \rvert} \sum_{m \in \tau(n)} D^*(m) - \mu_{i^*} \leq  -\delta  \right) \nonumber \\
&\leq e^{-2 n \theta^2} + e^{-2 \lvert \tau(n) \rvert \delta^2}
\label{arrivals_dominate_departures3}
\end{align}
where the last inequality follows from Hoeffding's inequality.

We finally note that by \eqref{cardinality_of_tau}, for any $q \in (0, 1 - \sigma )$, $\exists n_1$ such that $\forall n \geq n_1$, $\lvert \tau(n) \rvert \geq q n$.  Thus, $\forall n \geq \max \left\{ n_0, n_1, \frac{p + 1}{\delta - \theta} \right\}$
\begin{equation*}
\eqref{arrivals_dominate_departures3} \leq e^{-2 n \theta^2} + e^{-2 q n \delta^2} \leq 2 e^{- c_3 n}
\end{equation*}
for appropriately chosen positive constant $c_3$, which is independent of $p$ or $n$.  Noting that for appropriately chosen positive constants $c_4$ and $c_5$, $\max \left\{ n_0, n_1, \frac{p + 1}{\delta - \theta} \right\} \leq c_4 + c_5 p$ gives the lemma.
\end{IEEEproof}



\ifCLASSOPTIONcaptionsoff
  \newpage
\fi

\end{document}